\documentclass[a4paper, onecolumn, noarxiv, accepted=2022-05-22]{quantumarticle}

\usepackage[utf8]{inputenc}
\usepackage{color}
\usepackage{bm}% bold math
\usepackage{xcolor}
\usepackage{amsmath}
\usepackage{amssymb}
\usepackage{graphicx}
\usepackage[numbers, sort&compress]{natbib}
\usepackage{braket}
\usepackage{psfrag}
\usepackage{tikz}
\usepackage[normalem]{ulem}
\usepackage{qcircuit}
\usepackage{ulem}
\usepackage{multirow}
\usepackage{algorithmic}
\usepackage[section]{algorithm}
\usepackage{hyperref}
\hypersetup{colorlinks=true, citecolor=blue, urlcolor=blue, linkcolor=blue}
\usepackage[caption=false]{subfig}
\allowdisplaybreaks

\newcommand{\qft}{\mathcal{F}}
\newcommand{\boldx}{\mathbf{x}}
\newcommand{\boldxp}{\mathbf{x}'}
\newcommand{\boldy}{\mathbf{y}}
\newcommand{\boldyp}{\mathbf{y}'}
\newcommand{\boldz}{\mathbf{z}}
\newcommand{\boldS}{\mathbf{S}}
\newcommand{\bolddelta}{\bm{\delta}}

\newcommand\at[2]{\left.#1\right|_{#2}}
\newcommand{\ephase}{\epsilon_{\textsubscript{phase}}}
\newcommand{\identity}{\mathbb{I}}

\newtheorem{theorem}{Theorem}
\newtheorem{example}{Example}
\newtheorem{definition}{Definition}
\newcommand{\bigO}[1]{\mathcal{O}\left( #1 \right)}
\newcommand{\bigOt}[1]{\widetilde{\mathcal{O}}\left( #1 \right)}

\begin{document}

\title{Towards Quantum Advantage in Financial Market Risk using Quantum Gradient Algorithms}

\author{Nikitas Stamatopoulos}
\affiliation{Goldman, Sachs \& Co., New York, NY}

\author{Guglielmo Mazzola}
\affiliation{IBM Quantum, IBM Research -- Zurich}

\author{Stefan Woerner}
\affiliation{IBM Quantum, IBM Research -- Zurich}

\author{William J. Zeng}
\affiliation{Goldman, Sachs \& Co., New York, NY}

\begin{abstract}
We introduce a quantum algorithm to compute the market risk of financial
derivatives.
Previous work has shown that quantum amplitude estimation can accelerate derivative
pricing quadratically in the target error and we extend this to
a quadratic error scaling advantage in market risk computation.
We show that employing quantum gradient estimation algorithms can deliver a further
quadratic advantage in the number of the associated market sensitivities,
usually called \emph{greeks}.
By numerically simulating the quantum gradient estimation algorithms on
financial derivatives of practical interest, we demonstrate that not only can
we successfully estimate the greeks in the examples studied, but that the
resource requirements can be significantly lower in practice than what
is expected by theoretical complexity bounds.
This additional advantage in the computation of financial market risk lowers
the estimated logical clock rate required for
financial quantum advantage from
Chakrabarti et al.~\href{http://dx.doi.org/10.22331/q-2021-06-01-463}{[Quantum 5, 463 (2021)]} by a factor of $\sim 7$,
from 50MHz to 7MHz, even for a modest number of greeks by industry standards (four).
Moreover, we show that if we have access to enough resources, the quantum algorithm can be parallelized across up to 60 QPUs, in which case the logical clock rate of each device required to achieve the same overall runtime as the serial execution would be $\sim 100$kHz.
Throughout this work, we summarize and compare several different combinations of
quantum and classical approaches that could be used for computing the market risk of financial derivatives.
\end{abstract}

\maketitle

\section{Introduction}
Recently, quantum algorithms have been proposed to accelerate the pricing and risk analysis of financial
derivatives~\cite{rebentrost2018quantum, Woerner_2019, egger2019credit, Stamatopoulos_2020, chakrabarti2021threshold}.
These algorithms use quantum amplitude estimation to achieve quadratic advantage
compared to the
classical Monte Carlo methods that are used in practice for most
computationally expensive pricing.
Let $\epsilon_p$ be the error in pricing.
The quantum advantage stems from the runtime of a
classical Monte Carlo simulation
scaling as $\bigO{1/\epsilon_p^2}$, while the quantum algorithms have scaling
$\bigO{1/\epsilon_p}$~\cite{montanaro2015quantum}.

A related and important financial application is the computation of the sensitivity of
derivative prices to model and market parameters. This amounts to computing
gradients of the derivative price with respect to input parameters.
A primary business use of calculating these gradients is to enable
hedging of the market risk that arises from exposure to derivative
contracts. Hedging this risk is of critical importance to financial
firms~\cite{Hull}.
In some cases, gradients can be computed
analytically, e.g. when the derivative price has an analytical form.
In this work, we consider computing gradients where there is not a
closed form for the price and where the classical comparison is to Monte
Carlo simulation.
Gradients of financial derivatives are typically called~\emph{greeks}, as these quantities are commonly labeled using Greek alphabet letters.
For $k$ greeks, i.e. for a $k$-dimensional gradient,
\emph{classical finite difference} methods compute the price at multiple points
in each parameter dimension for a scaling of $\bigO{k/\epsilon_p^2}$.

One approach to quantum acceleration of greek computation is to construct a finite difference approximation of the financial derivative's price on a quantum computer and perform amplitude estimation
on that quantity instead of its price.
In Sec.~\ref{sec:SQG} we show that this approach, called the~\emph{semi-classical}
method in~\cite{gilyen2019optimizing}, scales as $\bigO{k/\epsilon}$, where $\epsilon$ is the target error for the gradients.
However, one can further improve the scaling in the number of greeks $k$,
using quantum algorithms for
computing the gradients~\cite{Jordan_2005, gilyen2019optimizing}.
A quantum algorithm to compute the gradient's components in superposition was
originally introduced by Jordan in Ref.~\cite{Jordan_2005} and more recently
revisited by Gily\'en, Arunachalam, and Wiebe (GAW)~\cite{gilyen2019optimizing}.

Gily\'en et al.~\cite{gilyen2019optimizing} perform a rigorous analysis
of Jordan's original quantum gradient algorithm~\cite{Jordan_2005} to show  that in general it scales as $\bigO{\sqrt{k}/\epsilon^2}$.
They then generalize that algorithm to arbitrary higher order $m$ and show
that an order $m=\log(\sqrt{k}/\epsilon)$ quantum gradient
algorithm has runtime of $\bigO{\sqrt{k}/\epsilon}$ for a specific class of smooth functions.
For the sake of simplicity, in the rest of the manuscript we call this result
as \emph{GAW quantum gradient} algorithm, from the name of the authors of Ref.~\cite{gilyen2019optimizing}.
Quantum gradient algorithms have also been previously studied as a subroutine for
accelerated convex optimization~\cite{Chakrabarti2020quantumalgorithms}.
Since, in this optimization context,
subtleties connected with the error's scaling and  parameters setting were found to be
important, it is timely to put forward a comprehensive study on the advantage
that could be found in financial applications.

In this work, we apply the algorithmic framework
from~\cite{gilyen2019optimizing} to compute financial greeks and study several
varieties of quantum gradient algorithms.
We use the quantum pricing method of
Chakrabarti et al.~\cite{chakrabarti2021threshold}
as a subroutine to the GAW algorithm to numerically estimate the resource
requirements of computing the greeks of two types of option contracts (a) a European call option, which we use as a
benchmark to establish the validity of the algorithm and estimate the
corresponding resource requirements, and (b) a path-dependent basket option
whose pricing profile is representative of typical financial derivative
contracts of practical interest.

Then, we introduce a second-order accurate quantum gradient algorithm, the $m=1$
version of~\cite{gilyen2019optimizing}, for which we can give an explicit and compact
quantum implementation in the financial derivative case, and which does not rely on block encoding or Hamiltonian simulation.
We use this \emph{Simulation-Free Quantum Gradient} (SFQG) method to compute the greeks of a path-dependent derivative and find that it is significantly cheaper to construct than the Hamiltonian-based method.
Additionally, we show that we can improve the overall performance of quantum gradient estimation algorithms by employing a maximum likelihood (MLE) method to extract the most likely estimate of the gradients with concrete confidence intervals.
With these tools, we calculate that quantum advantage for calculating risk may be achievable with quantum
computers whose clock rates are 7 times slower than
that required for pricing itself.
We discuss these implications in more detail
in Sec.~\ref{sec:advantage_estimates}.

Finally, we perform a comparison between quantum, classical and semi-classical
gradient estimation algorithms in the context of financial derivatives,
summarized in Table~\ref{tbl:numerical_basket}.

We highlight the new contributions in this work:
\begin{itemize}
\item We numerically study the quantum gradient estimation
algorithms from \cite{Jordan_2005, gilyen2019optimizing} for functions of
practical interest to financial market risk and compare the observed oracular cost
to theoretical expectations. (Section~\ref{sec:resource-estimates})
\item We devise a method to construct a second-order accurate oracle for
quantum gradient estimation for functions computed using quantum amplitude
estimation that is cheaper in required resources compared to existing methods.
(Section~\ref{sec:sfqg})
\item We introduce a way to improve gradient estimation algorithms using
classical maximum likelihood estimation (MLE). (Section~\ref{sec:mle})
\item We propose a technique to employ automatic differentiation (AD) methods
on quantum computers which can enhance the quantum gradient estimation
performance in certain cases. (Appendix~\ref{app:auto_diff})
\item We update the resource estimates for quantum advantage in financial
derivative pricing from prior research. (Section~\ref{sec:advantage_estimates})
\end{itemize}

\subsection{Quantum Amplitude Estimation}
\label{sec:ae}
Our gradient calculation algorithms extend the
quantum accelerated method for derivative pricing that is based on quantum
amplitude estimation (QAE)~\cite{brassard2002quantum}. We review QAE here.
Let the $k$-dimensional vector
$\boldx$ denote the set of market data parameters of the derivative represented as a basis state $\ket{\boldx}$ and $f
(\boldx)$ its price, rescaled to satisfy $f(\boldx) \in [0, 1]$.
We assume the existence of a unitary operator $\mathcal{A}$ which produces
the state
\begin{equation}
	\label{eqn:A_operator}
    \mathcal{A} : \ket{\vec{0}}\ket{\boldx} \rightarrow \left(\sqrt{1-f
    (\boldx)
	}\ket{\psi_0(\boldx)}\ket{0} + \sqrt{f(\boldx)}\ket{\psi_1(\boldx)}\ket{1} \right) \ket{\boldx},
\end{equation}
where $\ket{\psi_0(\boldx)}$ and $\ket{\psi_1(\boldx)}$ are arbitrary, normalized quantum states.
The state $\ket{\boldx}$ representing the market data parameters acts only as input to $\mathcal{A}$ and we can ignore it for the remainder of this section.

Explicit constructions of $\mathcal{A}$ with resource estimates
for specific path dependent derivative instances are given
in Chakrabarti et al.~\cite{chakrabarti2021threshold}.
QAE estimates $f(\boldx)$ with repeated applications of the operator
$\mathcal{Q} = \mathcal{A}S_0\mathcal{A}^{\dagger}
S _{\psi_0}$, where $S_0 = \mathbb{I} - 2 \ket{\vec{0}}\bra{\vec{0}}$ and $S_{\psi_0} = \mathbb{I} - 2\ket{\psi_0(\boldx)}\ket{0}\bra{0}\bra{\psi_0(\boldx)}$.
The QAE algorithm relies on the fact that, by the construction of $\mathcal{Q}$, the state produced after the application of
$\mathcal{A}$ in Eq.~\eqref{eqn:A_operator} can be written in the
eigenbasis of $\mathcal{Q}$ \cite{brassard2002quantum}
\begin{equation}
	\label{eqn:A_operator_eigenbasis}
	\frac{-i}{\sqrt{2}}\left(e^{i \theta(\boldx)} \ket{\psi_{+}(\boldx)}
	- e^{-i \theta(\boldx)} \ket{\psi_{-}(\boldx)}\right) \equiv \ket{\Psi(\boldx)},
\end{equation}
where $\ket{\psi_{+}(\boldx)}$ and $\ket{\psi_{-}(\boldx)}$ are the
eigenvectors of $\mathcal{Q}$ with eigenvalues $e^{\pm 2 i\theta
(\boldx)}$, and $f(\boldx) = \sin^2(\theta(\boldx))$.
Let integer $m>0$ denote the bits of precision with which we want to estimate the amplitude.
Applying $H^{\otimes m}$ on an $m$-qubit register initialized at $\ket{0}_m$
and using the register to control different powers of $\mathcal{Q}$,
a phase kickback of the eigenvalues of $\mathcal{Q}$ is induced onto
the control register
\begin{equation}
	\label{eqn:Q_operator_eigenbasis}
	\frac{1}{\sqrt{M}}\sum_{j=0}^{M-1}\ket{j}\mathcal{Q}^{j}
	\ket{\Psi(\boldx)}=\frac{1}{\sqrt{M}}
	\sum_{j=0}^{M-1}e^{2ij\theta(\boldx)}\ket{j}\ket{\Psi_{+}(\boldx)}
	- \frac{1}{\sqrt{M}}\sum_{j=0}^{M-1}e^{-2ij\theta(\boldx)}\ket{j}
	\ket{\Psi_{-}(\boldx)},
\end{equation}
where $\ket{\Psi_{\pm}(\boldx)} = -ie^{\pm i\theta(\boldx)}
\ket{\psi_{\pm}(\boldx)}/\sqrt{2}$ and $M=2^m$.
Finally, applying an inverse Quantum Fourier Transform on the first register
and measuring in the computational basis gives an $m$-bit approximation of
either $\tilde{\theta}_{+}(\boldx)=M\theta(\boldx)/\pi$ or
$\tilde{\theta}_{-}(\boldx)=-M\theta(\boldx)/\pi$. This can then be classically
mapped to an estimate for $f(\boldx)$ using
\begin{equation}
	\label{eqn:ae_mapping}
    \tilde{f}(\boldx) = \sin^2\left(\frac{\pi\tilde{\theta}_{\pm}(\boldx)
    }{M}\right).
\end{equation}
In order to approximate $\tilde{f}(\boldx)$ within additive error
$\epsilon_{p}$ with high probability, QAE requires $M = \bigO{1/\epsilon_{p}}$
invocations to $\mathcal{Q}$.
We measure the query complexity of the QAE algorithm by counting the number of
calls to the oracles $\mathcal{A}$ and $\mathcal{A}^{\dagger}$
and this scales as $\bigO{2/\epsilon_{p}}$.
In the case of financial derivatives, the function $f$ is the expectation value of the payoff of a derivative
contract. We explore this context in the following section.

\section{Gradient Methods for Financial Derivatives}
In this section we describe several methods for computing the gradients of
financial derivatives.
The complexities of these approaches are summarized in Table~\ref{tab:complexities}.

\begin{table}[t!]
  \begin{center}
	  \resizebox{0.95\columnwidth}{!}{
    \begin{tabular}{c|c|c|c|c}
      Classical Finite Difference
        & CFD with CRN
        & Semi-classical
        & Jordan's
        & GAW Quantum Gradient \\
      \hline \hline
      $\bigO{k/\epsilon^3}$ & $\bigO{k/\epsilon^2}$
      & $\bigO{k/\epsilon}$
      & $\bigO{\sqrt{k}/\epsilon^2}$
      & $\bigO{\sqrt{k}/\epsilon}$ \\
    \end{tabular}
	  }
    \caption{Summary of the complexity scaling of the different market risk
    algorithms. Here $\epsilon$ is the absolute error in gradient estimation,
    and $k$ is the dimension of the gradient. For the CFD and semi-classical
    methods, which depend additionally on the discretization step $h$, we assume the optimal choices, which are derived in Section~\ref{sec:cfd}.}
    \label{tab:complexities}
  \end{center}
\end{table}

\subsection{Greeks}
\label{sec:greeks}
Let $\boldS^t \in \mathbb{R}_+^d$ be a vector of values for $d$ underlying
assets at time $t$. Let $(\boldS^1, ..., \boldS^T) = \bar{\omega} \in
\bar{\Omega}$ be a path of a discrete time multivariate stochastic process
describing the values of those assets in time when the current (today's) market values of these assets are $\boldS^0$. We use both notations for a path in the
text. The corresponding probability density function is denoted by
$\bar{p}(\boldS^0, \bar\omega)$. Let $g(\boldS^0, \bar \omega) =
g\left(\boldS^0, ..., \boldS^T\right) \in
\mathbb{R}$ be the discounted payoff of some derivative on those assets. To
price the derivative we calculate
\begin{align}
\label{eq:price}
\mathbb{E}(g) = \int_{\bar{\omega}\in\bar{\Omega}}\bar{p}(\boldS^0, \bar{\omega})g
(\boldS^0, \bar{\omega})d\bar{\omega}
\approx \sum_{\omega\in\Omega}p(\boldS^0, \omega)g(\boldS^0, \omega),
\end{align}
where we have removed the bar notation to indicate that we have switched from a
continuous to a discrete model of prices.
Gradients of this price are known as greeks.
\begin{example}[Delta]
Consider a single underlying ($d=1$). We define the gradient of the
underlying with respect to the spot price $S^0$ as
\begin{align}
\Delta = \frac{\partial\mathbb{E}(g)}{\partial S^0}.
\end{align}
\end{example}

Other commonly used greeks are gradients of the price with respect to the time
$T$ (\emph{theta}), the volatility of the underlying model for $S$
(\emph{vega}), the correlation between assets, or other parameters. In general
one calculates $\mathbb{E}(g, \boldx)$ for some model and/or market
parameters $\boldx$ and then wishes to compute the set
$\{\frac{\partial\mathbb{E}(g)}{\partial \theta_{i=1...k}}\}$.
Remarkably, $k$
can be on the order of hundreds or thousands in practical cases.
For this reason, a
$k$ scaling improvement represents an important advantage in this context.

\subsection{Classical Finite Difference}
\label{sec:cfd}
Let $f(\boldx) = \mathbb{E}(g, \boldx)$ be the pricing function for a fixed
payoff $g$. Classically, we can compute the
gradients of a function $f: \mathbb{R}^k \rightarrow \mathbb{R}$ using
finite-difference methods by sampling the function $f$ over a sufficiently
small region $h$ so that expanding $f$ to first order gives a good
approximation to
$f(\boldx) \approx f(\mathbf{a}) + (\boldx-\mathbf{a}) \cdot \nabla f$.
The simplest \emph{forward} finite-difference classical scheme to compute the gradients at a
point $\boldx_0$ requires $k+1$ evaluations of $f$, one at $\boldx_0$ and $k$
evaluations displaced from $\boldx_0$ by $h$ in each dimension.
The gradient is then approximated using $\partial f/\partial x_i \approx [f
(\boldx_0 + h\hat{e}_i) -f(\boldx_0)]/h + \mathcal{O}(h)$, where $\hat{e}_i$
is the $i$th normalized basis vector.
For a more accurate approximation, we can instead use a second-order
scheme which requires $2k$ function evaluations, and approximate the gradients
with
\begin{equation}
	\label{eqn:central_difference}
\partial f/\partial x_i \approx \frac{f(\boldx_0 + \frac{h}{2}\hat{e}_i) - f
(\boldx_0 - \frac{h}{2}\hat{e}_i)}{h} + \mathcal{O}(h^2).
\end{equation}

Now suppose that the function $f(x)$ is evaluated with a finite accuracy $\delta \geq 0$.
The associated error for the forward or central finite differences formulas reads
\begin{eqnarray}
\label{eqn:fd_error_delta}
\frac{f(x+h) - f(x-(p-1)h) + \mathcal{O}(\delta)}{p h} + \mathcal{O}(h^p) &=& \partial_x f(x) + \mathcal{O}(\delta / h + h^p),
\end{eqnarray}
where $p = 1$ for forward and $p=2$ for central finite differences.
Suppose now that we want to achieve an overall estimation error of $\epsilon > 0$,  i.e.
\begin{eqnarray}
\delta / h + h^p \leq \epsilon,
\end{eqnarray}
where we focus on the asymptotic scaling.
While the step size $h$ can be freely chosen, improving the accuracy $\delta$ is usually related to increasing computational costs.
Thus, we want to maximize $\delta$ by setting $h$ while still achieving the target accuracy $\epsilon$.
This leads to
\begin{eqnarray}
\delta = (\epsilon - h^p) h = \epsilon h - h^{p+1}.
\end{eqnarray}
Setting the first derivative of the right-hand-side with respect to $h$ to zero leads to
\begin{eqnarray}
h = (\epsilon / (p+1))^{(1/p)},
\end{eqnarray}
which leads to the optimal $\delta$ for a target $\epsilon$ given by
\begin{eqnarray}
\delta &=& \mathcal{O}(\epsilon^{1 + 1/p}).
\end{eqnarray}

When $f(x)$ is approximated by algorithms relying on sampling, the number of samples for a target approximation error scales as $\delta = \mathcal{O}(1/M^q)$, where $M$ denotes here the number of samples and $q$ depends on the convergence rate of the algorithm i.e., $q = 1/2$ for classical Monte Carlo simulation and $q = 1$ for QAE.
Then, combining everything together leads to
\begin{eqnarray}
\label{eqn:fd_complexity}
M = \mathcal{O}(  \epsilon^{-(1+1/p)/q} ).
\end{eqnarray}
Thus, using $p=2$ and $q=1/2$, we calculate that the complexity of computing $k$ greeks using a central-difference method with Monte Carlo is $\mathcal{O}(k/\epsilon^3)$.
We call this approach the~\emph{classical finite
difference} (CFD) method.

\subsection{Finite-Difference with Common Random Numbers}
\label{sec:common_random_numbers}
Another approach for computing greeks, when derivative pricing is done
classically with Monte Carlo, is to use the second-order central-difference
method of Eq.~\eqref{eqn:central_difference}, but to perform correlated
sampling by using the same random
numbers in the Monte Carlo evaluation of both $f(\boldx_0 +
\frac{h}{2}\hat{e}_i)$ and $f(\boldx_0 - \frac{h}{2}\hat{e}_i)$.
This way the statistical fluctuations present in both terms cancel out, effectively removing the overall error dependence on the
discretization step $h$.
The complexity of evaluating $k$ greeks in this
case is $\mathcal{O}(k/\epsilon^2)$, which is also classically optimal \cite{Glasserman1992SomeGA}.
We call this method the \emph{classical finite difference with common random
numbers} (CFD-CRN).

\subsection{Semi-classical Quantum Gradients}
\label{sec:SQG}
A straightforward approach to improve the finite-difference method using quantum computation
is to use a central-difference formula with quantum amplitude estimation algorithm for each pricing
step~\cite{rebentrost2018quantum, Stamatopoulos_2020, chakrabarti2021threshold}.
This \emph{semi-classical quantum gradient} (SQG) method~\cite{gilyen2019optimizing}
then scales as $\mathcal{O}(k/\epsilon^{1.5})$, which we get by substituting $p=2$ and $q=1$ in Eq.~\eqref{eqn:fd_complexity}.
This approach can be improved upon in a similar manner to using CRN for
classical finite difference. Instead of computing $\mathbb{E}\left(f(\boldx_0 +
\frac{h}{2}\hat{e}_i)\right)$ and $\mathbb{E}\left(f(\boldx_0 -
\frac{h}{2}\hat{e}_i)\right)$ separately using amplitude estimation, we
compute $\mathbb{E}\left((f(\boldx_0 + \frac{h}{2}\hat{e}_i) - f_{\omega}
(\boldx_0 - \frac{h}{2}\hat{e}_i))/h\right)$.
We can do this by computing the quantity $\ket{(f_{\omega}(\boldx_0 + \frac{h}{2}\hat{e}_i)
- f_{\omega}(\boldx_0 - \frac{h}{2}\hat{e}_i))/h}$ in a quantum register, where $f_{\omega}$ denotes
the payoff for each path $\omega$ (cf. Sec.~\ref{sec:greeks}) and then using
amplitude estimation on this quantity.
In this case the output of amplitude estimation is
\begin{equation}
\label{eqn:sqg_fd}
\sum_{\omega\in\Omega}p(\omega)\left(f_{\omega}(\boldx_0 + (h/2)\hat{e}_i) - f_{\omega}(\boldx_0 - (h/2)\hat{e}_i)\right)/h \approx \mathbb{E}(\partial_i f(\boldx)),
\end{equation}
giving us the expectation value of the finite difference approximation of the $i$th derivative.
The advantage of this method is that there are no separate statistical fluctuations associated individually with $f_{\omega}(\boldx_0 \pm (h/2)\hat{e}_i)$ and we recover the standard $\mathcal{O}(1/\epsilon)$ scaling of amplitude estimation.
For $k$ gradients, we then get an overall complexity of $\mathcal{O}(k/\epsilon)$.

\subsection{Quantum Gradients}
\label{sec:quantum_gradients}
There are other quantum approaches that use a quantum Fourier transform to
compute the gradient with an improved scaling in the dimension
$k$~\cite{Jordan_2005, gilyen2019optimizing}. These algorithms require access
to a fractional \emph{phase oracle} for the target
function $f:\mathbb{R}^k\mapsto \mathbb{R}$, in our case the pricing
function of Sec.~\ref{sec:ae}.
This oracle, given a point $\boldx \in \mathbb{R}^k$ and $S > 0$, performs
the operation
\begin{equation}
	\label{eqn:general_phase_oracle}
	O_{Sf} : \ket{\boldx} \rightarrow e^{2\pi i S f(\boldx)}\ket{\boldx}.
\end{equation}
In order to estimate the $k$-dimensional gradient of $f$ at point $\boldx_0$,
the oracle is evaluated over a uniform superposition of points $\bolddelta$ in
a sufficiently small $k$-dimensional hypercube $G_{\boldx_0}^k$ of edge length $l$ around
$\boldx_0$ where each dimension is discretized using $N$ points with $n=\log N$ qubits, such that
\begin{equation}
	\label{eqn:general_gradient}
	\frac{1}{\sqrt{N^k}}\sum_{\bolddelta \in G_{\boldx_0}^k}O_{Sf}\ket{\bolddelta}
	= \frac{1}{\sqrt{N^k}}\sum_{\bolddelta \in G_{\boldx_0}^k}
	e^{2\pi i S f(\mathbf{x_0}+\bolddelta)}\ket{\bolddelta}
	\approx e^{2\pi i S f(\boldx_0)}\frac{1}{\sqrt{N^k}}
	\sum_{\bolddelta \in G_{\boldx_0}^k}
	e^{2\pi i S \cdot \nabla f_{\boldx_0}\cdot\bolddelta}\ket{\bolddelta},
\end{equation}
assuming $f(\boldx_0 + \bolddelta) \approx f(\boldx_0) + \nabla f_{\boldx_0}
\cdot \bolddelta$ for $||\bolddelta|| \ll 1$.
As shown in \cite{Jordan_2005, gilyen2019optimizing}, choosing $S=N/l$ and applying the $k$-dimensional inverse Quantum Fourier Transform on the resulting
state, we measure in the
computational basis to get an approximation of $\nabla f_{\boldx_0}$ with
precision $\epsilon = \mathcal{O}(1/N)$ and high probability.
The complexity of these gradient estimation algorithms depends on the resources
required to construct the oracle of Eq.~\eqref{eqn:general_phase_oracle} and
the size of $G_{\boldx_0}^k$ required to make the approximation in
Eq.~\eqref{eqn:general_gradient} sufficiently accurate.
As we discuss in the next section in more detail, the computational cost of the gradient estimation method depends on the value of $S=N/l$.
Intuitively, the cost of the method increases as the desired precision of the gradient estimate increases (larger $N$), and the value of $l$ denoting the region we evaluate the function $f$ decreases.
While the value of $N$ is an input to the gradient estimation method controlling the desired precision, the required value of $l$ depends on the degree of non-linearity of the function $f$ in the region of evaluation around $\boldx_0$.
The more non-linear the function $f$ is around $\boldx_0$, the smaller $l$ we need to pick so that the function can be sufficiently well-approximated by the first order Taylor expansion used in Eq.~\eqref{eqn:general_gradient}.

\section{Higher-order methods for quantum gradients}
\label{sec:gilyen_method}
Gily\'en et al~\cite{gilyen2019optimizing} showed that in order to
estimate $k$ gradients of a function $f$ with accuracy $\epsilon$, the oracle in Eq.~\eqref{eqn:general_phase_oracle} has to be
evaluated $S = N/l \sim D_2\sqrt{k}/\epsilon^2$ times, where $D_2$ is an upper bound on the magnitude of the second-order derivatives of $f$ and $N,l$ as defined in Section~\ref{sec:quantum_gradients}.
While this gives a quadratically improved scaling in $k$, it does not improve on
the error scaling of the semi-classical method.
To improve also the scaling in
accuracy $\epsilon$, Ref.~\cite{gilyen2019optimizing} introduces higher-degree
central-difference schemes.

The $2m$-point central-difference approximation for the gradient of $f$ at
$\mathbf{0}$ is given by
\begin{equation}
	\label{eqn:2m_point_difference}
	\nabla f_{(2m)}(\mathbf{0}) = \sum_{\ell=-m}^{m}a_{\ell}^{(2m)}f(\ell
	\boldx),
\end{equation}
where the coefficients $a_{\ell}^{(2m)}$ depend on the choice of $m$ and for
uniform grid spacing we have $a_{\ell}^{(2m)}=-a_{-\ell}^{(2m)}$
\cite{Fornberg1988GenerationOF}.
A phase oracle for the general $2m$-point scheme
of Eq.~\eqref{eqn:2m_point_difference} can be constructed by composing
individual fractional phase oracles for each of the $2m$ terms, scaled by the
appropriate coefficient $a_{\ell}^{(2m)}$.
This leads to a family of algorithms with different phase oracles at different
orders $m$
\begin{equation}
	\label{eqn:m_phase_oracle}
	O_{Sf}^m : \ket{\boldx} \rightarrow e^{2\pi i S
	\sum_{\ell=-m}^{m}a_{\ell}^{(2m)}f(\ell
	\boldx)
	}\ket{\boldx},
\end{equation}
with $S=N/l$.
For example, the phase oracle for the two-point approximation
\begin{equation}
\label{eqn:gilyen_oracle}
O_f^1\ket{\boldx} = e^{2\pi i (f(\boldx)- f(\mathbf{-x}))/2} \ket{\boldx},
\end{equation}
can be constructed as the product of oracles $O_f^{+}\ket{\boldx} = e^{\pi i f
(\boldx)} \ket{\boldx}$ and $O_f^{-}\ket{\boldx} = e^{-\pi i f(-\boldx)} \ket{\boldx}$.

With these higher-order methods, they show that estimating the
gradients of a class of smooth functions\footnote{Also known as Gevrey class
$G^{\frac{1}{2}}$ functions \cite{ASENS_1918_3_35__129_0}} to accuracy
$\epsilon$ using Jordan's algorithm, can instead be done with $S = \mathcal{O}
(\sqrt{k}/\epsilon)$ phase oracle applications by picking a large enough value
for $m$ in the central-difference approximation used.
Therefore, for this family of smooth functions, gradient estimation using
high-order central-difference methods scales quadratically better in the
desired accuracy $\epsilon$ compared to the original Jordan's algorithm. This is
described formally in the following Theorem:

\begin{theorem}[5.4 from~\cite{gilyen2019optimizing}]
\label{thm:smoothness_conditions}
Let $\boldx \in \mathbb{R}^k, \epsilon < c \in \mathbb{R}_+$ be fixed constants
and suppose $f:\mathbb{R}^k\mapsto \mathbb{R}$ is analytic and satisfies the
following: for every $j\in\mathbb{N}$ and $\alpha\in [k]^j$
\begin{equation}
|\partial_{\alpha}f(\boldx)| \le c^j j^{\frac{j}{2}}.
\end{equation}
Using the GAW Algorithm and setting $m=\log(c\sqrt{k}/\epsilon)$ we can compute an $\epsilon$-approximate gradient
$\tilde{\nabla}f(\boldx) \in \mathbb{R}^k$ such that
\begin{equation}
\|\nabla f(\boldx) - \tilde{\nabla}f(\boldx) \|_{\infty} \le \epsilon,
\end{equation}
with probability at least $1-\delta$, using
$\bigOt{
\frac{c\sqrt{k}}{\epsilon}
\log\left(\frac{k}{\delta}\right)
}$ queries to
a probability or (fractional) phase oracle of $f$.
\end{theorem}

In addition to the desired $\mathcal{O}(\sqrt{k}/\epsilon)$ scaling, the complexity of the GAW algorithm in Theorem \ref{thm:smoothness_conditions} includes a factor of $\log(k/\delta)$ stemming from the fact that we need to extract the medians after the application of the quantum gradient estimation algorithm, with probability of $1-\delta$.
While in the following sections we focus on the $\mathcal{O}(\sqrt{k}/\epsilon)$ factor in order to establish the dominant scaling of the complexity with respect to $k$, in Sec.~\ref{sec:mle} we show how we can eliminate the $\log(k/\delta)$ factor by using classical maximum likelihood estimation.
This approach not only decreases the overall computational complexity of the algorithm, but also provides us with concrete confidence intervals for a given confidence level.

We summarize in Table~\ref{tab:complexities} the scaling in $k$ and $\epsilon$ of the algorithms discussed in this Section.

\subsection{Creating phase oracles from probability oracles}
\label{sec:prob_to_phase_oracle}

Because the function whose gradients we would like to compute is only
accessible
through a \emph{probability oracle} in the form of Eq.~\eqref{eqn:A_operator}, we need to create a corresponding \emph{phase oracle} of Eq.~\eqref{eqn:general_phase_oracle} in order to use the quantum gradient method described in this section.
To do so, we can use the
block encoding technique from Ref.~\cite{gilyen2019optimizing}:

\begin{definition}[4.4 from~\cite{gilyen2019optimizing}]
	\label{def:block-encoding}
	Suppose that $A$ is an operator on a Hilbert space $\mathcal{H}$, then we say that the unitary $U$ acting on $\mathcal{H}_{\text{aux}} \otimes \mathcal{H}$ is a block-encoding of $A$ if
	\begin{equation*}
		A = (\bra{\vec{0}} \otimes \identity)U (\ket{\vec{0}} \otimes \identity).
	\end{equation*}
\end{definition}
Intuitively, the block-encoding $U$ is a unitary whose top-left block contains $A$:
\begin{equation}
	U = \begin{bmatrix}
			A & \text{  .  } \\
			. & \text{  .  }
		\end{bmatrix}.
\end{equation}
For the probability oracle $\mathcal{A}$ of Eq.~\eqref{eqn:A_operator}, the authors of Ref.~\cite{gilyen2019optimizing} observe that
\begin{equation}
	\label{eqn:block_encoding_A}
	(\bra{\vec{0}} \otimes \identity)(\mathcal{A}^{\dagger} (Z \otimes \identity) \mathcal{A})(\ket{\vec{0}} \otimes \identity) = \text{diag}(1-2f(\boldx)),
\end{equation}
and from Definition~\ref{def:block-encoding}, $U \equiv \mathcal{A}^{\dagger} (Z \otimes \identity) \mathcal{A}$ is a block-encoding of a diagonal matrix $H$ with diagonal entries $(1-2f(\boldx))$.
With access to this block-encoding, the Hamiltonian simulation method from \cite{low2019hamiltonian, gilyen2019quantum} allows us to implement an $\ephase$-approximation of the unitary $e^{itH}$, through repeated applications of $U$ and $U^{\dagger}$, which scales as $\mathcal{O}(|t| + \ln(1/\ephase))$.
Therefore, for appropriately chosen values of $t$, we can use block-encoding and Hamiltonian simulation to produce an $\ephase$-approximation of the phase oracle $O_{Sf}$ from Eq.~\eqref{eqn:general_phase_oracle}.
Note that while the block-encoding of Eq.~\eqref{eqn:block_encoding_A} allows us to create a phase oracle with phase of $1-2f(\boldx)$ instead of $f(\boldx)$ in Eq.~\eqref{eqn:general_phase_oracle}, the first term ends up as a global phase which can be ignored, and the factor of $-2$ can be absorbed in the factor $t$ of the Hamiltonian simulation.

While references \cite{low2019hamiltonian, gilyen2019quantum} describe how to perform Hamiltonian simulation and the resources required to realize the necessary unitary evolution, the methods presented therein require post-selection, which needs to be factored in for an end-to-end resource estimation.
On the other hand, Ref. \cite{martyn2021efficient} introduces a coherent Hamiltonian simulation method which does not require post-selection and instead succeeds with arbitrarily high probability $1-\delta$, scaling as $\mathcal{O}(|t| + \ln(1/\ephase) + \ln(1/\delta))$.
For a target $t$, approximation error $\ephase$ and $\delta = 2\ephase$, this coherent Hamiltonian simulation algorithm queries $U$ and its inverse a total number of times
\begin{equation}
	N_U(t, \ephase, \beta) = 2 \left\lfloor \frac{1}{2}r\left( \frac{e|t|}{2 \beta}, \frac{5\ephase}{24} \right) \right\rfloor +\gamma\left(\frac{\ephase}{3}, 1-\beta \right) + 1,
\end{equation}
where
\begin{itemize}
	\item $\beta \in (0, 1)$ is a user-chosen parameter,
	\item $r(\tau, \epsilon) = |\tau|e^{W(\ln(1/\epsilon)/|\tau|)}$, where $W(x)$ is the Lambert-$W$ function,
	\item $\gamma(\epsilon, \Delta) = 2 \cdot \left\lceil \max \left(\frac{e}{\Delta}\sqrt{W\left(\frac{8}{\pi\epsilon^2}\right) W\left(\frac{512}{e^2\pi \epsilon^2}\right)}, \sqrt{2}W\left(\frac{8 \sqrt{2}}{\sqrt{\pi}\Delta \epsilon}\sqrt{W\left(\frac{8}{\pi \epsilon^2}\right)} \right) \right) \right\rceil$ + 1.
\end{itemize}

Using this method, the total number of oracle (Eq.~\eqref{eqn:A_operator}) calls required to construct an $\ephase$-approximation of the $m$-order phase oracle of Eq.~\eqref{eqn:m_phase_oracle} is then given by
\begin{equation}
	\label{eqn:No}
	N_{o} = \sum_{\ell=-m}^{m} N_U\left(2\pi\frac{N}{l}|a_{\ell}^{(2m)}|, \ephase, \beta \right).
\end{equation}
The optimal value of $\beta$ depends on the target approximation error $\ephase$.
For the cases studied in this manuscript, we find that $\beta=0.5$ is the optimal choice and we fix that value whenever we use Eq.~\eqref{eqn:No}.

\section{Resource Estimation of Quantum Gradient Methods}
\label{sec:resource-estimates}
In this section, we perform an asymptotic resource estimation for the gradient
methods described previously.
We choose representative parameters from the financial domain for gradient
estimation problems, targeting $k=1000$ greeks and an approximation error of $\epsilon = 10^{-3}$.
For the resource estimation of the GAW method, we assume that the smoothness conditions of
Theorem~\ref{thm:smoothness_conditions}, with smoothness parameter $c=1$, apply
for the problem at hand and that the Hamiltonian simulation phase error is
$\ephase = 10^{-4}$.
Here we choose $c=1$ in order to estimate the possible usefulness of the algorithm in a best-case scenario of smoothness from Theorem~\ref{thm:smoothness_conditions}.
While we treat the phase error $\ephase$ as a free parameter at this point, our particular choice of $10^{-4}$ is motivated by numerical simulations which show this to be a good choice.
We discuss the numerical simulations and the impact of the phase error in more detail in subsequent sections.
Then, we set the finite-difference approximation degree to
\begin{equation}
\label{eqn:_gaw_m}
m = \log(c\sqrt{k}/\epsilon)
\end{equation}
and the spacing parameter to\footnote{See proof of Theorem 5.4
from~\cite{gilyen2019optimizing}}
\begin{equation}
\label{eqn:gaw_spacing}
l^{-1} = 9cm\sqrt{k}\left(81\times 8 \times 42\pi
cm\sqrt{k}/\epsilon\right)^{1/(2m)}.
\end{equation}
Using the proofs from~\cite{gilyen2019optimizing} we can estimate the number of
oracle calls that will be required to achieve the target error.
Notice that, in these
estimates, we consider only the asymptotic scaling (assuming remaining constant
factors are
1) and assume that $\ephase$ is sufficiently small to have no impact on
performance. This results in $6.3 \times 10^7$ oracle calls.

We can compare this asymptotically to the performance of both classical finite
difference (with common random numbers and without) and semi-classical quantum finite difference. These results are
summarized in Table~\ref{tab:resources}.
We choose the optimal values of the discretization step $h$ as calculated in Section~\ref{sec:cfd} to minimize the gradient estimation error $\epsilon$, and pick $\epsilon=10^{-3}$ for these benchmarks.

\begin{table}[h!]
  \begin{center}
    \begin{tabular}{c|c|c|c}
      Classical Finite Difference & CFD with CRN & Semi-classical & GAW Quantum
       Gradient \\
      \hline \hline
      $10^{12}$ & $10^{9}$ & $10^6$ & $10^7$ \\
    \end{tabular}
    \caption{Table of asymptotic oracle calls required to compute quantum
    gradients for financially relevant greeks. Parameters for this benchmark
    are described in the text. This asymptotic analysis indicates that quantum gradient method has
    potential to outperform classical finite difference methods but is on par with semi-classical methods.
    }
    \label{tab:resources}
  \end{center}
\end{table}

In Table \ref{tab:resources}, we notice that using the parameters from the proofs in \cite{gilyen2019optimizing}, the GAW algorithm fails to deliver an advantage compared to the semi-classical method for $k$ as high as $10^3$.
However, while this analysis gives us an idea of how the different methods compare in theory, the estimates are based on asymptotic bounds with parameters which may be loose in practice and are highly dependent on the smoothness of the functions considered.
In the following section we study the performance of the GAW method numerically
 on small examples that are representative of some practical cases in finance.

\subsection{GAW Numerical estimates}
\label{sec:gaw_numerical}
The GAW method described in Sec.~\ref{sec:gilyen_method} gives a quantum
algorithm for gradient estimation which scales as $\mathcal{O}(\sqrt{k}/\epsilon)$ when the target function satisfies the conditions of Theorem~\ref{thm:smoothness_conditions}.
However, in most relevant financial models of interest, we do not have access
to closed-form solutions that we can examine to check whether they satisfy the
smoothness conditions required of the theorem.
As such, in this section we numerically examine the behavior of the high-order
methods of the GAW gradient estimation algorithm for two financial use cases:
(a) A simple (\emph{vanilla}) European call option for which we have an
analytical closed form solution and can benchmark the performance of the
algorithm against the exact gradients, and (b) a path-dependent basket option
with a~\emph{knock-in} feature which has no known analytical solution and is in
practice classically evaluated using Monte Carlo methods.
In particular, we examine the central-difference approximation order $m$ and
spacing $l$ required for adequately precise gradient estimation and compare it
to the theoretical values of Eq.~\eqref{eqn:_gaw_m} and Eq.~\eqref{eqn:gaw_spacing} respectively.
We focus on these two parameters because they determine the overall complexity of the algorithm, which we can then compare to the other methods of Table~\ref{tab:resources}.

Because the resulting quantum circuits are prohibitively wide and deep for
numerical simulation in practice, we adopt the following practical method to
emulate the algorithm's performance:
to estimate $k$ greeks using $n$ bits of precision we initialize a
$k*2^n$-dimensional array with the amplitudes of Eq.~\eqref{eqn:m_phase_oracle} computed classically for the chosen derivative order $m$.
We then perform a $k*2^n$-dimensional classical inverse Fourier transform of the array to get the resulting probability distribution which is the output of the GAW algorithm before measurement.
In order to account for the phase error $\epsilon_{\textsubscript{phase}} > 0$ from the Hamiltonian simulation, we add a random number to each encoded phase in Eq.~\eqref{eqn:m_phase_oracle}, uniformly picked from the interval $[-\epsilon_{\textsubscript{phase}}, \epsilon_{\textsubscript{phase}}]$.

\subsubsection{Vanilla Options}
\label{sec:vanillas}
The simplest example in derivative pricing is a European call option whose price depends on the performance of a single asset at a pre-determined future time (the \emph{expiration} date), relative to a reference price (the \emph{strike}).
In the Black-Scholes-Merton model \cite{BlackScholes}, where the asset
undergoes Geometric Brownian Motion (GBM), a call option on a non dividend-paying asset has a closed form solution given by
\begin{equation}
	\label{eqn:vanilla_price}
	C = SN(d_1) - Ke^{-rT}N(d_2), \quad  \text{ with } d_1=\frac{\ln(S/K) + (r
	+ \sigma^2/2)T}{\sigma\sqrt{T}},\; d_2=d_1 - \sigma\sqrt{T},
\end{equation}
where $S$ is the asset price today, $K$ the strike of the option, $r$ is a risk-free rate of return, $\sigma$ the annualized volatility of the asset, $T$ is the time until the option's expiration date and $N(x)$ denotes the CDF of the standard normal distribution.
We test the quantum gradient estimation algorithm for the four greeks of this
option/model: $\textit{delta} = \partial{C}/{\partial S}$, $\textit{rho} =
\partial{C}/\partial{r}$, $\textit{vega} = \partial{C}/{\partial{\sigma}}$, and $\textit{theta} = \partial{C}/{\partial T}$.
We numerically simulate the GAW algorithm for increasing $k$ (the number of
greeks we compute simultaneously) and central-difference approximation order $m
 \in [1, 4]$.
In each case we search for the largest value of the spacing
$l$ in for which the algorithm produces an estimate $\epsilon$-close to the exact value with probability $\ge 85\%$ for each greek.
We target a gradient error of $\epsilon = 2 \times 10^{-2}$ which requires
$n=\lceil\log(1/\epsilon)\rceil = 6$ qubits in each dimension.~\footnote{Our
numerical simulations scale exponentially in this parameter as are emulating
the quantum circuit on these qubits. This limits us to this size.}

\begin{figure}[t]
  \centering
  \includegraphics[width=0.85\linewidth]{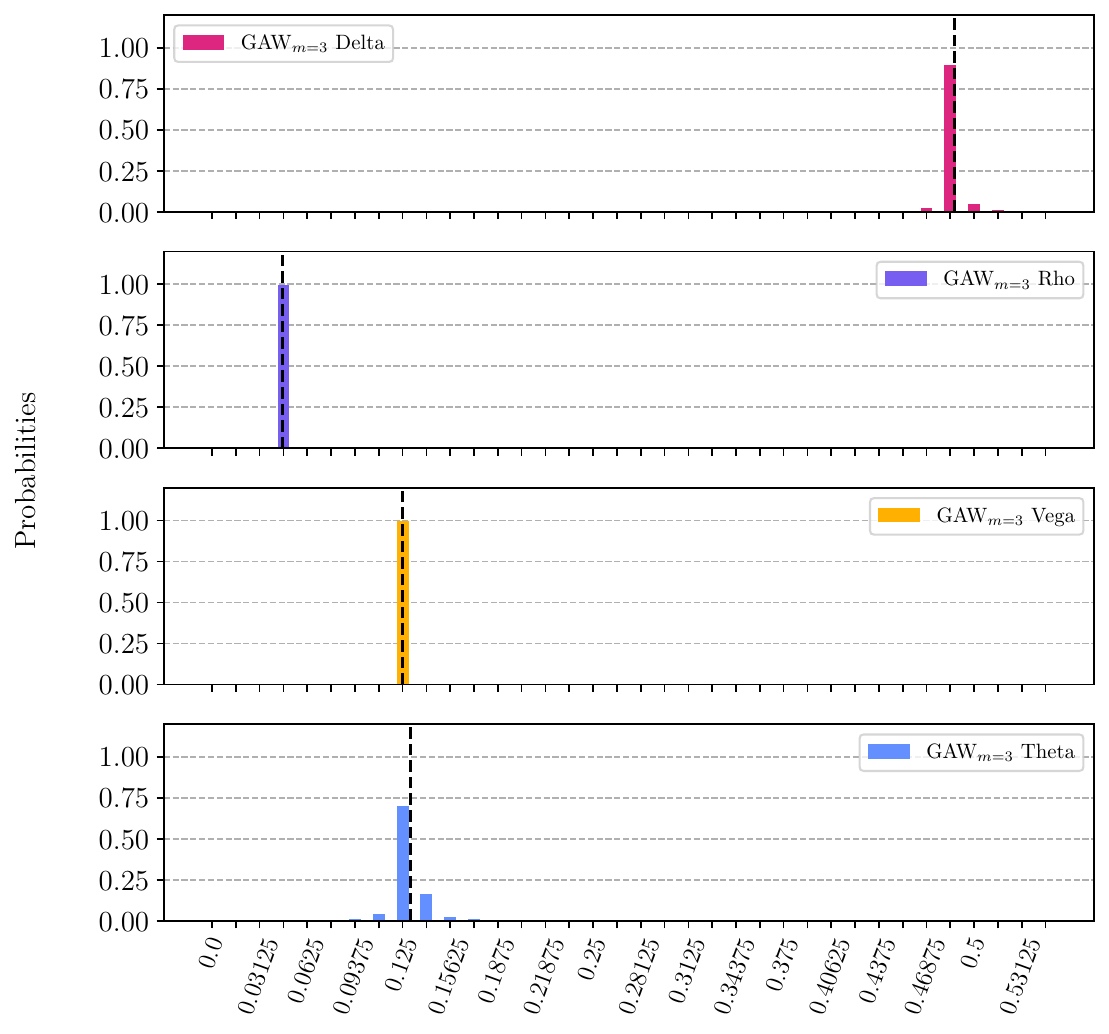}
  \caption{The probability distribution resulting from the numerical simulation of the GAW quantum gradient estimation algorithm with $m=3$  and $l=0.58$ for the four greeks (delta, rho, vega, theta) of the vanilla option of Eq.~\eqref{eqn:vanilla_price}, and the corresponding exact values shown as dashed vertical lines. Measurement of the registers corresponding to each greek will result in a value at most $\epsilon = 1/N = 2\times 10^{-2}$ away from the exact value with probability $\ge 85\%$.}
  \label{fig:vanilla_histograms}
\end{figure}

For each value of $k$, we then compare what we numerically find as the optimal
values for $(m, l)$ --- that minimize the total number of oracle calls $N_{o}$
from Eq.~\eqref{eqn:No} while maintaining a success
probability of $\ge 85\%$ --- to those used in the proof of
Theorem~\ref{thm:smoothness_conditions}, given by Eq.~\eqref{eqn:_gaw_m} and
Eq.~\eqref{eqn:gaw_spacing}, assuming the smoothest possible parameter $c=1$.
We set the approximation error from the Hamiltonian simulation to
$\ephase=10^{-4}$ and include it as an error source in our numerical simulations.
The gradients of the vanilla option in Eq.~\eqref{eqn:vanilla_price} are
evaluated at the point $(S, r, \sigma, T) = (99.5, 1\%, 20\%, 0.1)$, with
$K=100$, chosen so that the parameter values are reasonably realistic from a
finance point of view, but at the same time probing a domain where the function
is as non-linear as possible.\footnote{Where the function is (approximately)
linear then simple finite difference methods perform well already.}
In Table~\ref{tbl:numerical_vanilla} we show the results of the numerical simulation and the corresponding theoretical estimates from Theorem~\ref{thm:smoothness_conditions} and in Fig.~\ref{fig:vanilla_histograms} we show the resulting probability distribution from the quantum gradient estimation algorithm for each greek, for the case $k=4$ in Table~\ref{tbl:numerical_vanilla}.
In order to simulate the algorithm for increasing values of $k$, we pick $k$ out of the four parameters $S, r, \sigma, T$ with respect to which we compute the gradients and fix the values of the remaining $4-k$ parameters.
For the the cases $k \in [2,3]$ where we have a choice of which parameters we fix, we have verified that the simulation results in Table~\ref{tbl:numerical_vanilla} are qualitatively independent of the choice.

\begin{table}[h!]{
\bgroup
\setlength{\tabcolsep}{12pt}
 \def\arraystretch{1.5}%
  \begin{tabular}{l||c|c|c||c|c|c||}
       & \multicolumn{3}{c||}{Numerical}  & \multicolumn{3}{c||}{Theoretical}\\
       \hline
    	$k$ & $m$ & $l$ & $N_o$ & $m$ & $l$ & $N_o$  \\
	   \hline
	    2 ($delta, rho$) & 1 & 0.65 & 1976 & 5 & 0.0028 & 570592 \\
	    3 ($delta, rho, vega$) & 3 & 0.65 & 4664 & 5 & 0.0022 & 712008 \\
	    4 ($delta, rho, vega, theta$) & 3 & 0.58 & 4904 & 5 & 0.0019 & 833296

  \end{tabular}
  \egroup
  }
  \caption{In this table we show (a) numerical estimates of the query
  complexity $N_o$ (Eq.~\eqref{eqn:No}) and parameter values $(m, l)$ required by the GAW gradient
  estimation algorithm in order to estimate $k$ greeks for the vanilla option
  of Eq.~\eqref{eqn:vanilla_price} within $\epsilon = 2\times 10^{-2}$ with
  probability $\ge 85\%$ and (b) the corresponding values used in the proof of Theorem~\ref{thm:smoothness_conditions}. We notice that for a vanilla option, we can reduce the central-difference order and increase the spacing by more than two orders of magnitude compared to the theoretical values. As such, the query complexity we estimate numerically is $\sim 200$ times smaller than the theoretical.}
\label{tbl:numerical_vanilla}
\end{table}

Interestingly, we notice from Table~\ref{tbl:numerical_vanilla} that the query complexity required to estimate the gradients of the vanilla option with high probability in practice, is orders of magnitude smaller than what is expected from the parameters used in proof of Theorem~\ref{thm:smoothness_conditions}.
Because the types of vanilla options explored in this section are primarily a
motivating example of relevance to finance that are simpler to analyze, we now
turn our attention to more complex derivatives for which gradient estimation is required for business use in practice.

\subsubsection{Path-dependent Basket Options}
\label{sec:basket_option}
In the previous section, we numerically examined the performance of the GAW
algorithm for vanilla options for which we have analytical solutions and we can benchmark the algorithm's performance compared to the exact gradients of the model.
We now perform a similar analysis to a path-dependent option on multiple
underlying assets, which has no known closed form solution, and compare the
query complexity from the numerical simulation to the theoretical complexity from Theorem~\ref{thm:smoothness_conditions} as well as to that of the SQG method (Sec.~\ref{sec:SQG}) which calculates gradients using finite-difference using values estimated using QAE.
Similarly to the previous section, we set the approximation error from the Hamiltonian simulation to
$\ephase=10^{-4}$.

The option we consider in this section is defined on three underlying assets undergoing GBM with volatilities $\sigma_1=20\%, \sigma_2=20\%, \sigma_3=10\%$ and spot prices $\vec{S}(t=0) = (S_1(t=0), S_2(t=0), S_3(t=0)) = (2.0, 2.0, 2.0)$.
The risk free rate is set to $r=1\%$ and the option expires in $T=3$ years.
The weighted sum of the asset prices $\vec{w} \cdot \vec{S}(t)$ with weights $\vec{w} = (w_1, w_2, w_3) = (0.5, 0.3, 0.2)$ is observed on five days $t^B = [T/5 * i]$ for $i \in [1, 5]$ across the duration of the contract and the option's payoff is given by
\begin{equation}
	\label{eqn:basket_payoff}
	f(\vec{S}(t), K, B) =
	\begin{cases}
		\max(\vec{w} \cdot \vec{S}(T) - K, 0), & \text{if}\  \vec{w} \cdot \vec{S}(t) > B \text{ for any}\ t \in t^B \\
		0, & \text{otherwise,}
	\end{cases}
\end{equation}
where we set the strike $K = 1.0$ and $B = 2.5$.
This option is said to have a \emph{knock-in} feature, because it only pays off (i.e. is \emph{knocked-in}) if the observed weighted sum exceeds the knock-in barrier $B$ at any of the pre-defined times before the contract's expiration.
We simulate the GAW algorithm to estimate four gradients of this option contract's price $V$: the three deltas $(\partial V/\partial S_i)$ and the vega with respect to the first asset $\partial V/\partial \sigma_1$.
Because this option does not have a closed-form solution, we price the option using classical Monte Carlo with $10^6$ paths and use finite-difference to compute the expected gradients which we use as benchmarks for the quantum algorithm.
The quantum algorithm is simulated using $n=4$ qubits for each gradient register, which sets the target error of the algorithm to $\epsilon \le 1/2^4 = 0.0625$.
Using $m=1$ and $l=0.25$ the simulated GAW algorithm gives us an $\epsilon$-close estimate with probability $\ge 85\%$ for each greek.
The SQG method requires $1/\epsilon$ calls to the $\mathcal{Q}$ operator of QAE, and each $\mathcal{Q}$ includes two calls to the unitary $\mathcal{A}$ of Eq.~\eqref{eqn:A_operator} and its inverse, therefore the complexity for $k$ greeks is $2k/\epsilon$.
Because with the SQG method we need to compute the payoff twice in order to construct the finite difference (Eq.~\eqref{eqn:sqg_fd}), the $\mathcal{A}$ operator will be approximately twice as large as the regular pricing oracle.
In order to compare the query complexity more accurately with the other quantum methods we thus include a factor of two in the complexity, for a total of $4k/\epsilon$.

\begin{figure}[th]
  \centering
  \includegraphics[width=0.85\linewidth]{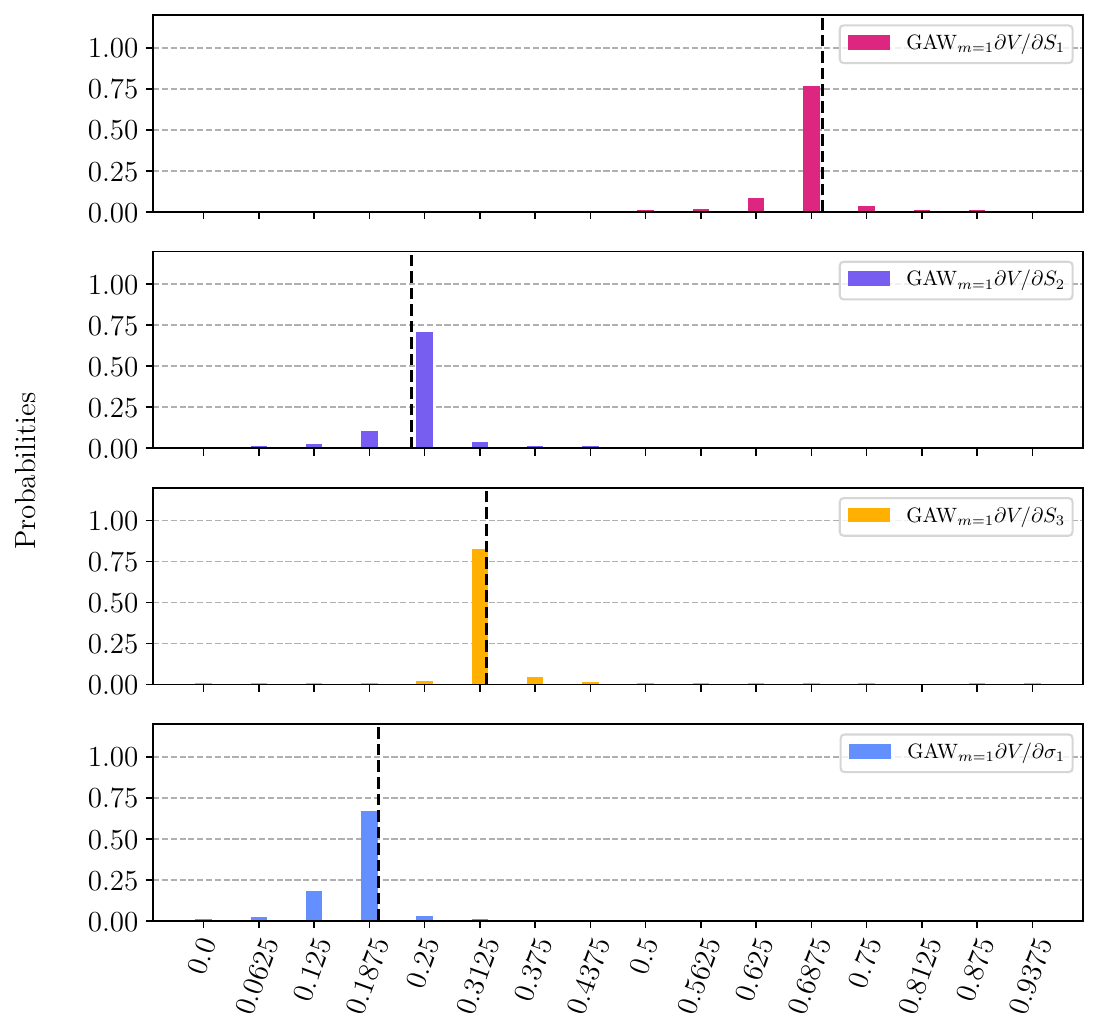}
  \caption{The probability distribution resulting from the numerical simulation of the GAW quantum gradient estimation algorithm with $m=1$ and $l=0.25$ for the four greeks of the path-dependent basket option of Sec.~\ref{sec:basket_option}, along with the values estimated using classical Monte Carlo (MC) with $10^6$ paths (vertical dashed lines). Measurement gives us estimates for each greek with error $\epsilon \le 0.0625$ with probability $\ge 85\%$.}
  \label{fig:basket_histograms}
\end{figure}

In Table~\ref{tbl:numerical_basket} we show the query complexity and parameters
from the numerical simulation of the GAW method for this path-dependent basket
option, along with (a) the asymptotic estimates from
Theorem~\ref{thm:smoothness_conditions}, (b) the query complexity of the SQG method and (c) the query complexity of the CFD and CFD-CRN methods, all for the same target approximation error.
For the CFD and CFD-CRN methods, the reported query complexity is the total number of Monte Carlo paths required for the evaluation of the $k$ greeks within the target approximation error $\epsilon$ with probability $\ge 85\%$, computed numerically.
The table also includes the parameters and resources required for the same calculation using the \emph{Simulation-Free Quantum Gradient} (SFQG) algorithm described in the next section.
The resulting probability distribution for all greeks along with the MC-estimated ``true" values is shown in Fig.~\ref{fig:basket_histograms}.

\section{Simulation-Free Quantum Gradient Method}
\label{sec:sfqg}

In this section, we construct a second-order accurate
quantum gradient algorithm, corresponding to $m=1$ in Eq.~\eqref{eqn:2m_point_difference}.
This is higher order than Jordan's algorithm,
which has unfavorable scaling, but lower order than required from the analysis
in Gily\'en et al. to guarantee the $\mathcal{O}(\sqrt{k}/\epsilon)$ scaling.
A benefit is that, in the case of derivative pricing, we are able
to give an explicit construction of the phase oracle without needing
to appeal to block encoding or Hamiltonian simulation.
We call this the \emph{Simulation-Free Quantum Gradient} (SFQG) method.

To describe this construction, we first show how to turn the derivative pricing setting of Sec.~\ref{sec:ae} into a first-order phase oracle $O_{Sf}$ and then how to build a second-order extension
to construct $O^1_{Sf}$ as defined in Eq.~\eqref{eqn:m_phase_oracle}.
We then simulate this algorithm to estimate the greeks of the basket option from Sec.~\ref{sec:basket_option}, allowing us to compare its performance to the corresponding second-order accurate GAW method.

\subsection{First-Order Pricing Phase Oracle}
\label{sec:first_order_phase_oracle}
In order to apply the quantum gradient algorithms to estimate the gradient of a function $f(\boldx)$, we need to construct a phase
oracle of the form of Eq.~\eqref{eqn:general_phase_oracle}. We
evaluate this oracle on a superposition of points $\boldx$ in a
$k$-dimensional hypercube of edge length $l$, $G^k=[-l/2, l/2]^k$ around
$\boldx_0$, where each dimension is discretized using $N$ points with $n=\log N$ qubits and $l$ chosen small enough so that $f(\boldx)$ is
approximately linear in that region.
For simplicity, we let $\boldx_0=\mathbf{0}$ as gradient estimation at other points can be achieved by trivially redefining the function $f$.
Evaluating the $\mathcal{A}$ operator of Eq.~\eqref{eqn:A_operator} on all points $\boldx$ in superposition we get
\begin{eqnarray}
	\label{eqn:A_superposition}
	\frac{1}{\sqrt{N^k}}\sum_{\boldx}\ket{\boldx}\ket{0}_{q+1}
	&\xrightarrow{\mathcal{A}}&
							  \frac{1}{\sqrt{N^k}}\sum_{\boldx}\ket{\boldx}
							  \left(\sqrt{1-f(\boldx)}\ket{\psi_0}_q\ket{0} +
							  \sqrt{f(\boldx)}\ket{\psi_1}_q\ket{1} \right)
							  \nonumber \\
				              &=& \frac{1}{\sqrt{N^k}}\sum_{\boldx}\ket{\boldx}
				               \left(\ket{\Psi_{+}(\boldx)} - \ket{\Psi_{-}
				               (\boldx)}\right)
\end{eqnarray}
where $\ket{\Psi_{\pm}(\boldx)} = -ie^{\pm i\theta(\boldx)}\ket{\psi_{\pm}
(\boldx)}/\sqrt{2}$ and $f(\boldx)=\sin^2(\theta(\boldx))$, similarly to
Eq.~\eqref{eqn:Q_operator_eigenbasis}.
Now, define the Grover operator $\mathcal{Q} =
\mathcal{A}S_0\mathcal{A}^{\dagger}
S_{\psi_0}$, where $S_0 = \mathbb{I}^{\otimes nk} \otimes
\left(\mathbb{I} - 2 |0\rangle_{q+1}\langle 0|_{q+1}\right)$ and
$S_{\psi_0} = \mathbb{I}^{\otimes nk} \otimes \left(\mathbb{I} -
2|\psi_0\rangle_q|0\rangle\langle 0|\langle\psi_0|_q\right)$.
Let $D$ be an upper bound on $|(\partial\theta(\boldx)/\partial x_i)|$  for all
$i \in [1, k]$ and apply the $\mathcal{Q}$ operator $\pi S$ times to the state in
Eq.~\eqref{eqn:A_superposition} with $S=N/Dl$ to get \footnote{Because the states $\ket{\Psi_{\pm}(\boldx)}$ in Eq.\eqref{eqn:Q_phase_kickback} depend on $\boldx$, we must be careful to make sure they do not interfere with the phase kicked back to the $\ket{\boldx}$ register.
As we discuss in Appendix~\ref{app:phase_kickback} the structure of the $\mathcal{A}$ operator used in derivative pricing allows the phase kickback to take place correctly.}
\begin{equation}
	\label{eqn:Q_phase_kickback}
  	\mathcal{Q}^{\pi S}\mathcal{A} :
  	\frac{1}{\sqrt{N^k}}\sum_{\boldx}\ket{\boldx}\ket{0}_{q+1} \rightarrow \frac{1}{\sqrt{N^k}}\sum_{\boldx}e^{2\pi
   	iS\theta(\boldx)}\ket{\boldx} \ket{\Psi_{+}(\boldx)} -
    	\frac{1}{\sqrt{N^k}}\sum_{\boldx}e^{-2\pi iS\theta(\boldx)
    	}\ket{\boldx}\ket{\Psi_{-}(\boldx)}.
\end{equation}
Note that the application of $O_{S\theta} \equiv \mathcal{Q}^{\pi S}\mathcal{A}$ in
Eq.~\eqref{eqn:Q_phase_kickback} is close to the phase oracle we
are looking for in quantum gradient estimation, but
(i) for $\theta$ rather than $f$, and
(ii) with a superposition over the positive and negative phases that we seek.

An inverse Quantum Fourier Transform $\left(\qft^{-1}_n\right)^{\otimes k}$ on
the first register then gives us an estimate of the derivatives of $\theta
(\boldx)$ at $\boldx_0$
\begin{equation}
	\label{eqn:measured_gradients}
  	\Ket{\frac{N}{D}\frac{\partial \theta}{\partial
  	x_1}}\Ket{\frac{N}{D}\frac{\partial \theta}{\partial x_2}} \dots
   	\Ket{\frac{N}{D}\frac{\partial \theta}{\partial x_k}} \quad \text{or} \quad
    	 \Ket{-\frac{N}{D}\frac{\partial \theta}{\partial
     	 x_1}}\Ket{-\frac{N}{D}\frac{\partial \theta}{\partial x_2}} \dots
      	 \Ket{-\frac{N}{D}\frac{\partial \theta}{\partial x_k}}.
\end{equation}

This method gives us an oracle for $\theta$ and not for $f$ directly.
However, because we know that $\theta$ and $f$ are related through $f(\boldx)=\sin^2(\theta(\boldx))$,
we can compute the derivatives of $\theta$ and then use the chain rule to get the gradients of $f$:
\begin{equation}
	\label{eqn:chain_rule}
	\at{\pm\frac{\partial f}{\partial x_i}}{\boldx_0} =
	\at{\pm\frac{\partial \theta}{\partial x_i}}{\boldx_0} \times \sin\left(2\theta(\boldx_0)\right).
\end{equation}
This requires knowledge of $\theta(\boldx_0)$ which can be calculated
separately through standard QAE via the operator of Eq.~\eqref{eqn:A_operator}.

The positive and negative cases can be distinguished by adding a dummy
dimension to $f$ with a gradient that has a known sign. For example one could
transform the function $\left(f:\mathbb{R}^k\mapsto\mathbb{R}\right)\mapsto
\left(f+0.5x_{k+1}:\mathbb{R}^{k+1}\mapsto\mathbb{R}\right)$. Inspecting the sign
of the $k+1$-th derivative tells us if we are in the positive or negative case.

Because in this setting the value of $\pi S = \pi N/Dl$ indicates the number of times we need to invoke the $\mathcal{Q}$ operator in Eq.~\eqref{eqn:Q_phase_kickback}, it must be expressible as an integer.
Since $N = 2^n$ is an integer, this can be achieved by picking a value of $D$ such that $\pi/Dl$ is also an integer.
After we apply the $\mathcal{Q}$ operator the resulting integral number of times, the measured gradients in Eq.~\eqref{eqn:measured_gradients} are classically multiplied by the choice of $D$ in order to recover the correct magnitude of the gradients.

\subsection{Second-Order Pricing Phase Oracle}
\label{sec:second_order_exact_phase_oracle}
\begin{figure}[th]
  \centering
  \includegraphics[width=0.85\linewidth]{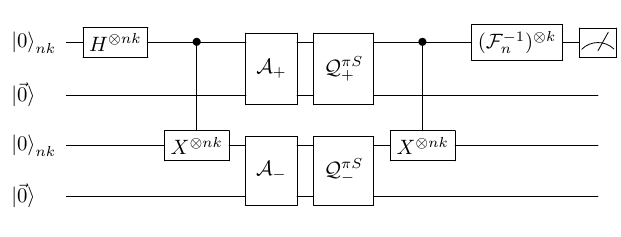}
  \caption{Circuit diagram of the SFQG method used to generate the state of Eq.~\eqref{eqn:sfqg_oracle_pp}. We first apply Hadamard gates to $k$ registers of $n$ qubits each to generate the superposition of points $\sum_{\boldx}\ket{\boldx}$. We then use CNOT gates to create a copy of each $\ket{\boldx}$ which allows us to evaluate the oracles $O_{S\theta}^+ \equiv \mathcal{Q}_+^{\pi S}\mathcal{A}_+$ and $O_{S\theta}^- \equiv \mathcal{Q}_-^{\pi S}\mathcal{A}_-$ in parallel at the cost of extra qubits. After uncomputing the copies, an $n$-dimensional inverse Quantum Fourier Transform in each of the initial $k$ registers gives us the state of Eq.~\eqref{eqn:sfqg_oracle_pp} which we then measure to estimate the $k$ gradients. }
  \label{fig:sfqg_circuit}
\end{figure}

The simplest high-order extension we can do is to construct an oracle which
encodes the two-point approximation shown in Eq.~\eqref{eqn:gilyen_oracle},
which corresponds to $m=1$ in Eq.~\eqref{eqn:2m_point_difference}.
Because in the QAE setting we are computing the gradients of $\theta(\boldx) =
\sin^{-1}\sqrt{f(\boldx)}$ instead of $f(\boldx)$, we need to construct an
oracle which performs
\begin{equation}
\label{eqn:sfqg_oracle}
O^1_{S\theta}\ket{\boldx} = e^{2\pi i S(\theta(\boldx)- \theta(-\boldx))/2}
\ket{\boldx},
\end{equation}
where $\theta(-\boldx)=\sin^{-1}\sqrt{f(-\boldx)}$.
In \cite{gilyen2019optimizing}, the authors suggest that
high-order oracles of this form can be constructed as the product of two
separate oracles with opposite phases.
The same idea can be used to construct the oracle of Eq.~\eqref{eqn:sfqg_oracle} using appropriately defined oracles.
Let $\mathcal{A}_+$ label the $\mathcal{A}$ operator of Eq.~\eqref{eqn:A_operator} and define operator $\mathcal{A}_-$ which acts as a probability oracle for the value $f(-\boldx)$

\begin{equation}
	\label{eqn:A_minus_operator}
    \mathcal{A}_- : \ket{\vec{0}}\ket{\boldx} \rightarrow \left(\sqrt{1-f
    (-\boldx)
	}\ket{\psi_0(-\boldx)}\ket{0} + \sqrt{f(-\boldx)}\ket{\psi_1(-\boldx)}\ket{1} \right) \ket{\boldx},
\end{equation}
as well as corresponding Grover operators $\mathcal{Q}_+ = \mathcal{A}_+S_0\mathcal{A}_+^{\dagger}S_{\psi_0}$ and $\mathcal{Q}_- = \mathcal{A}_-S_0\mathcal{A}_-^{\dagger}S_{\psi_0}$.
From Eq.~\eqref{eqn:Q_phase_kickback} we see that the product of the oracles $O_{S\theta}^+ \equiv \mathcal{Q}_+^{\pi S}\mathcal{A}_+$ and $O_{S\theta}^- \equiv \mathcal{Q}_-^{\pi S}\mathcal{A}_-$ generates the state
\begin{align}
	\label{eqn:sfqg_oracle_pp}
	& \frac{1}{\sqrt{N^k}}\sum_{\boldx}e^{2\pi iS(\theta(\boldx)
	- \theta(-\boldx))}\ket{\boldx} \ket{\psi_1(\boldx)}
	+ \frac{1}{\sqrt{N^k}}\sum_{\boldx}e^{-2\pi iS(\theta(\boldx)
	- \theta(-\boldx))}\ket{\boldx} \ket{\psi_2(\boldx)} \nonumber \\
	& + \frac{1}{\sqrt{N^k}}\sum_{\boldx}e^{2\pi iS(\theta(\boldx)
	+ \theta(-\boldx))}\ket{\boldx} \ket{\psi_3(\boldx)}
	+ \frac{1}{\sqrt{N^k}}\sum_{\boldx}e^{-2\pi iS(\theta(\boldx) + \theta
	(-\boldx))}\ket{\boldx} \ket{\psi_4(\boldx)}
\end{align}
where $\psi_{i}$ denotes products of eigenstates of $\mathcal{Q}_+$ and
$\mathcal{Q}_-$ which can then be ignored for the rest of the algorithm.
While the first two terms contain the appropriate phase kickback (up to the sign) for the
two-point approximation method of Eq.~\eqref{eqn:sfqg_oracle} with combined probability of $50\%$, the last
two terms encode a phase proportional to $\theta(\boldx) + \theta(-\boldx) =
\theta(\boldx_0) + \mathcal{O}(\partial^2\theta(\boldx)/\partial\boldx^2)$,
which create a probability peak around zero instead of the gradient, with similar combined $50\%$ probability.
By adding a dummy variable to the function as described at the end of Sec.~\ref{sec:first_order_phase_oracle}, we can distinguish which eigenstate we are in by measuring the gradient of the dummy variable after applying the inverse Quantum Fourier Transform.
Because we know the gradient with respect to the dummy variable by construction, measuring the positive (negative) gradient of the dummy variable means we are measuring the positive (negative) gradient with respect to the other variables.
Otherwise, if we measure zero in the dummy variable register, we ignore that measurement.
This means that additional post-processing is required for this method, and $50\%$ of the shots are discarded.
The total number of shots required depends on the desired accuracy of the estimation and this choice is discussed in more detail in Sec.~\ref{sec:mle}.
A circuit diagram of the SFQG method is shown in Fig.~\ref{fig:sfqg_circuit}.
Note that the operators $\mathcal{Q}_+$ and $\mathcal{Q}_-$ need to be applied to separate registers in order to generate the correct phases for Eq.~\eqref{eqn:sfqg_oracle_pp}.

Using the operator to construct the state of Eq.~\eqref{eqn:sfqg_oracle_pp}, we numerically simulate the SFQG algorithm to compute the greeks of the basket option of Sec.~\ref{sec:basket_option}.
In this case we use the algorithm to compute the gradients $\partial \theta/\partial S_1$, $\partial \theta/\partial S_2$, $\partial \theta/\partial S_3$, $\partial \theta/\partial \sigma_1$, where $\theta=\sin^{-1}\sqrt{V}$, and $V$ is the option price, from which we can then estimate the option's greeks using Eq.~\eqref{eqn:chain_rule}, assuming we have already priced the contract.
Similarly to the simulation of the GAW algorithm in Sec.~\ref{sec:basket_option}, we search for the parameter value $l$ which estimates the gradients within $\epsilon \le 0.0625$ with probability of success $\ge 85\%$, using $n=4$ qubits for each gradient register.
Here we ignore the probability of obtaining the states we discard in Eq.~\eqref{eqn:sfqg_oracle_pp} in post-processing.
Intuitively, we can think of the effective oracular cost as being twice what we compute from this simulation given that we have $50\%$ probability of failure, but as we show in Sec.~\ref{sec:mle} there is a more efficient way of combining measurement results to obtain estimates and confidence intervals for the gradients.
The complexity of the algorithm in terms of the number of serial invocations to the $\mathcal{A}$ operator of Eq.~\eqref{eqn:A_operator} required to construct the state of Eq.~\eqref{eqn:sfqg_oracle_pp} is then $\pi S=\pi N/l$, where $N=2^n$.

\begin{figure}[th]
  \centering
  \includegraphics[width=0.85\linewidth]{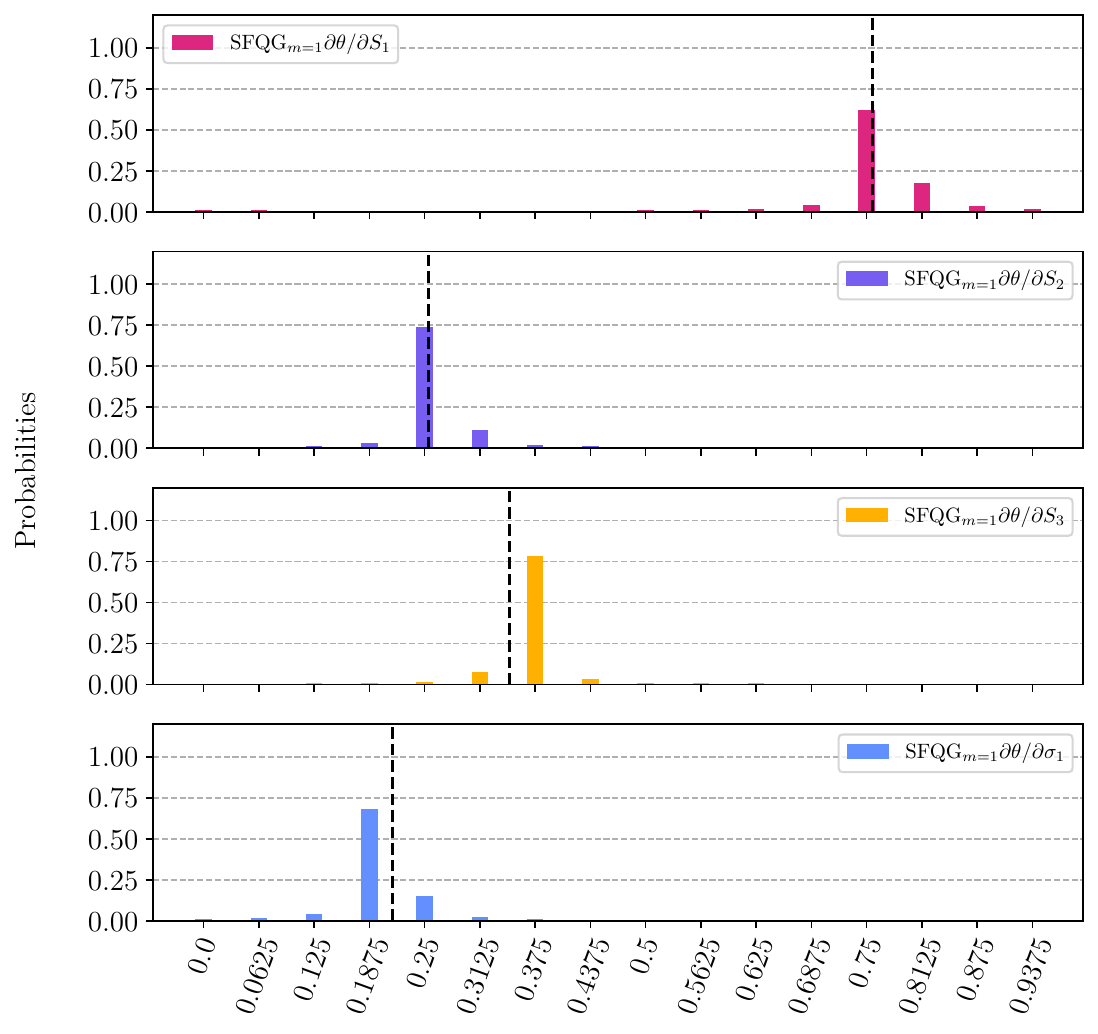}
  \caption{The probability distribution resulting from the numerical simulation of the second-order accurate, Simulation-Free Quantum Gradient (SFQG) algorithm described in Sec.~\ref{sec:sfqg} with $m=1$ and $l=0.25$ for the four greeks of the path-dependent basket option of Sec.~\ref{sec:basket_option}, along with the exact values estimated using classical Monte Carlo simulation (vertical dashed lines). This algorithm allows us to estimate the gradients of the function $\theta=\sin^{-1}\sqrt{V}$, where $V$ is the option price, from which we can then estimate the option's greeks using Eq.~\eqref{eqn:chain_rule}. Similarly to Fig.~\ref{fig:basket_histograms}, measurement gives us estimates for each greek within error $\epsilon \le 0.0625$ with probability $\ge 85\%$.}
  \label{fig:basket_histograms_theta}
\end{figure}

The probability distribution in the non-discarded states for all four gradients after the application of the SFQG algorithm is shown in Fig.~\ref{fig:basket_histograms_theta}, and the parameter values and query complexity of the algorithm is shown in Table \ref{tbl:numerical_basket} along with all the other quantum, classical and semi-classical methods studied in this manuscript.
The classical method complexity is estimated by computing the greeks with finite-difference using 3000 Monte Carlo simulations of the basket option price and searching for the number of Monte Carlo paths which give the same target error $\epsilon \le 0.0625$ with probability $\ge 85\%$ as the quantum and semi-classical methods.

\begin{table}[h!]{
\bgroup
\setlength{\tabcolsep}{10pt}
\def\arraystretch{1.4}%
  \begin{tabular}{c|r||c|c|c|}
       & & $N_o$  & m  &  l \\
       \hline
	   \parbox[t]{2mm}{\multirow{4}{*}{\rotatebox[origin=c]{90}{\textbf{Quantum}}}} & Simulation-Free (SFQG)* & 201 & 1 & 0.25 \\
	   \cline{2-5}
	   & Semi-classical (SQG) & 256 & n/a & n/a  \\
	   \cline{2-5}
	   & GAW (Numerical) & 1600 & 1 & 0.25 \\
	   \cline{2-5}
	   & GAW (Theoretical) & 201,528 & 4 & 0.0018 \\
	   \hline
	   \hline
	   \def\arraystretch{2.2}%
	   \begin{tabular}{l}
	   \parbox[t]{2mm}{\multirow{2}{*}{\rotatebox[origin=c]{90}{\textbf{Classical}}}}\end{tabular} & Finite-Difference w/ CRN (CFD-CRN) & 32,000 & n/a & n/a  \\
	   \cline{2-5}
	   & \def\arraystretch{2.2}%
	   \begin{tabular}{l} Finite-Difference (CFD)\end{tabular} & 400,000 & n/a & n/a \\

	  \hline

  \end{tabular}
  \egroup
  }
  \caption{Comparison between (a) the numerical estimates of the query
   complexity $N_o$ (Eq.~\eqref{eqn:No}) and parameter values $(m, l)$ required by the GAW gradient
   estimation algorithm in order to estimate $k$ greeks for the basket option
   of Eq.~\eqref{eqn:basket_payoff} within $\epsilon \le 0.0625$ with
   probability $\ge 85\%$, (b) the corresponding theoretical values used in the proof of
   Theorem~\ref{thm:smoothness_conditions}, (c) the parameters and resources
   required for the same calculation using the Simulation-Free Quantum Gradient
   (SFQG) algorithm described in Sec.~\ref{sec:sfqg}, (d) the query complexity
   of an SQG method for the same target accuracy and confidence interval and (e)
   the total number of simulated classical Monte Carlo paths required for the same
   accuracy and confidence interval using CFD and CFD-CRN methods. We
   numerically find that for this path-dependent basket option, the $m=1$ GAW method can estimate the four greeks within $\epsilon$ with
   probability $\ge 85\%$, with $\sim 125$ times smaller query complexity implied by the
   proof of Theorem~\ref{thm:smoothness_conditions} and 20 times smaller than the best
   finite-difference-based classical method (CFD-CRN). (* For the SFQG method, we can think of the effective oracular cost as being twice what is reported in this table given that we have $50\%$ probability of failure, but as we show in Sec.~\ref{sec:mle} there is a more efficient way of combining measurement results to obtain estimates and confidence intervals.)}
\label{tbl:numerical_basket}
\end{table}

\section{Quantum Gradient Estimation using Maximum Likelihood Estimation}
\label{sec:mle}
The success of the quantum gradient estimation algorithms studied so far in this manuscript, as well as those in \cite{Jordan_2005, gilyen2019optimizing} is judged by their ability to estimate the gradient within $\epsilon$ with high probability.
In \cite{gilyen2019optimizing} the authors suggest repeating the gradient estimation algorithm $\mathcal{O}(\log(k/\rho))$ times for $k$ gradients and taking the median to get the estimates within error $\epsilon$ with probability at least $(1-\rho)$.
However, this approach has two main drawbacks: a) In practice we would like to be able to characterize the correctness of the output with precise confidence intervals, and b) the output can only be one of the $N$ possible values of the discretized hypercube $G_{\boldx_0}^{k}$ of Eq.~\eqref{eqn:general_gradient}.

Recently, approaches using Maximum Likelihood Estimation (MLE) have been proposed to address these issues for amplitude estimation algorithms \cite{suzuki2020amplitude, tanaka2020amplitude, grinko2021iterative}, where the quantum circuits are sampled more than once, and the results are classically post-processed with MLE.
In this section we show that we can apply the MLE method from \cite{grinko2021iterative} to the quantum gradient estimation algorithms to enhance the final estimate.

Given a probability distribution $p$ with unknown parameter $g$ and data ${x_i}$ with $i=1,\dots, M$ sampled from it, MLE is used to obtain an estimate $\hat{g}$ for $g$.
This is done by maximizing the log-likelihood $\log L$

\begin{equation}
\label{eqn:log_likelihood}
\hat{g} = \arg\underset{g'}{\max}\log L(g') = \arg\underset{g'}{\max}\log \left( \prod_{i=1}^{M} p(x_i|g')\right)= \arg\underset{g'}{\max} \left( \sum_{i=1}^{M} \log p(x_i|g')\right),
\end{equation}
 which measures how likely it is to measure the data ${x_i}$ if $g'$ is the true parameter.
The general quantum gradient estimation algorithm from Sec.~\ref{sec:quantum_gradients} requires a region $l$ where the function whose gradients we are computing is approximately linear.
In this case, the probability distribution after the inverse Quantum Fourier Transform is applied to Eq.~\eqref{eqn:general_gradient} is the same as that of phase estimation and is given by \cite{brassard2002quantum}

\begin{equation}
    \label{eqn:ae_distribution}
    p(x) = \frac{\sin^2\left(N \Delta \pi \right)}{N^2 \sin^2\left(\Delta \pi \right)},
\end{equation}
where $\Delta = (x-g')$ and $N$ is the number of possible measurements.

Confidence intervals for the MLE estimate $\hat{g}$ can be derived using the likelihood ratio (LR) \cite{KochParameter}.
Following the analysis in the supplementary information of \cite{grinko2021iterative}, the confidence interval at the $(1-\alpha)$ confidence level is the value $g'$ satisfying $\left\{ g' \in [0,1] : \log L(g') \ge \log L(\hat{g}) - q_{\chi_1^2}(1-\alpha)/2 \right\}$, where $q_{\chi_1^2}$ denotes the $(1-\alpha)$ quantile of the $\chi^2$ distribution.

\begin{figure}[ht]
\subfloat[]{\label{fig:mle_distribution_a}\includegraphics[width=.49\linewidth]{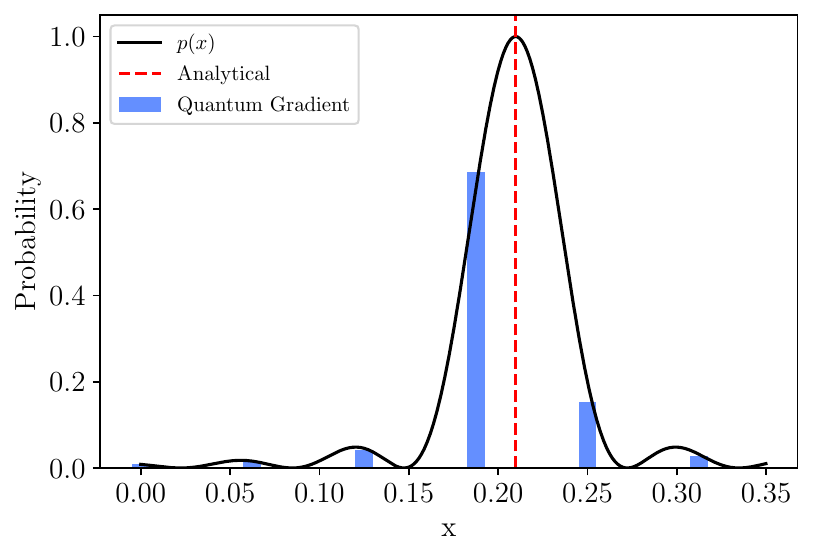} }%
%\qquad
\subfloat[]{\label{fig:mle_distribution_b}\includegraphics[width=.49\linewidth]{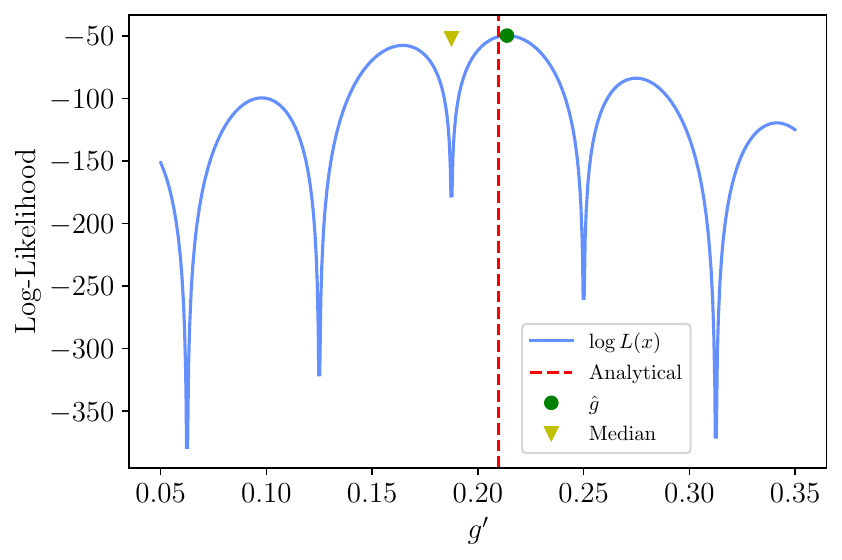} }%
\caption{a) The discrete probability distribution before measurement of Vega ($\partial \theta/\partial \sigma$) for the basket option from Fig.~\ref{fig:basket_histograms_theta} (blue bars) is fitted to the theoretical distribution $p(x)$ of Eq.\eqref{eqn:ae_distribution} (black line). The maximum of the distribution occurs at the exact analytical value (dashed red line). b) The global maximum of the log-likelihood $\log L(g')$ (green dot) gives us a better estimate of the true value (dashed red line) than the most likely result if we sample from the probability distribution and take the median (yellow triangle). The log-likelihood plot was produced by sampling the quantum gradient estimation circuit 30 times and allows us to estimate this greek to within $\epsilon \le 2 \times 10^{-3}$ with confidence level $1-\alpha=0.68$. When using the SFQG method we have $50\%$ probability of discarding a measurement, and the expected number of shots for this target accuracy is thus 60.}%
\label{fig:mle_distribution}
\end{figure}

In Fig.~\ref{fig:mle_distribution} we show how MLE can be used to improve the estimate of Vega ($\partial \theta/\partial \sigma$) for the basket option from Fig.~\ref{fig:basket_histograms_theta} calculated with the simulation-free quantum gradient method.
As discussed in Sec.~\ref{sec:second_order_exact_phase_oracle}, the SFQG method generates a probability peak of $50\%$ around the gradient up to the sign and another peak of $50\%$ around zero which we need to ignore.
Because we can distinguish between these cases by adding a dummy variable to the function with a known gradient, we can consider only the case where we measure the gradient, and double the number of required shots to take into account the discarded measurements.
First, in Fig.~\ref{fig:mle_distribution_a} we show that the final probability distribution after the application of the quantum gradient estimation algorithm does indeed fit Eq.~\eqref{eqn:ae_distribution}, and thus the QAE with MLE results from \cite{grinko2021iterative} can be used here too.
For clarity, we only plot the measurement outcomes in the interval which contains most of the probability mass.
In Fig.~\ref{fig:mle_distribution_b} we plot the log-likelihood $\log L(g')$ as a function of $g'$ across the same interval when we sample the quantum gradient circuit that produces the probability distribution in Fig.~\ref{fig:mle_distribution_a} 30 times.
The MLE estimate $\hat{g}$ is the global maximum of the function (green dot), which is very close to the true value $g$ (dashed red line) and significantly better than the median of the final probability distribution (yellow arrow).
In practice, the SFQG method requires double the number of shots (60) to account for the discarded shots due to the extra terms of Eq.~\eqref{eqn:sfqg_oracle_pp}.

We note that one significant advantage of the MLE method is that because the final estimate is not constrained to be one of the $N$ possible discrete values, we can decrease the value of $N$ at the cost of increasing the number of samples we take, thus lowering the overall width and depth of the quantum circuit.
For instance, while the distance between the $N$ possible discrete values in Fig.~\ref{fig:mle_distribution_a} is $0.0625$, the error $\epsilon$ in the MLE estimate in Fig.~\ref{fig:mle_distribution_b} is $\epsilon=4\times 10^{-3}$.

While the MLE post-processing requires additional classical compute cost, calculating the MLE in this setting is done by maximizing a concave function over a one-dimensional compact interval, which we ignore in the overall complexity analysis.
For more details on this process, we refer the reader to the supplementary information of \cite{grinko2021iterative}.

\section{Updated Estimates for Quantum Advantage}
\label{sec:advantage_estimates}
The resource estimation of Chakrabarti et al.~\cite{chakrabarti2021threshold} established that
quantum advantage for derivative pricing with respect to classical Monte Carlo
methods could require a quantum processor that can execute $\sim 10^7$ T-gates per second at a code distance that can support $\sim 10^{10}$ logical operations.
More specifically, pricing the autocallable contract studied in Ref.~\cite{chakrabarti2021threshold} using the reparameterization technique introduced therein to within $\epsilon \le 2\times 10^{-3}$ with confidence level $1-\alpha=0.68$ requires a T-depth of $5 \times 10^{7}$.
As such, in order to match the classical Monte Carlo pricing time estimated as 1 second, a logical quantum clock rate of $50$MHz would be needed.
We use the same reparameterization technique from Ref.~\cite{chakrabarti2021threshold}
to construct oracle size estimates, and estimate the T-depth
of the SFQG quantum gradient circuit required to compute the four greeks of the basket option of
Sec.~\ref{sec:basket_option} and Table \ref{tbl:numerical_basket}.
We find that in order to calculate the greeks to within the same $\epsilon \le 2\times 10^{-3}$ and confidence level $1-\alpha=0.68$, we can use the SFQG gradient estimation method with the parameters of Fig.~\ref{fig:basket_histograms_theta} and the maximum likelihood estimation (MLE) method from Sec.~\ref{sec:mle} using 60 shots (see Fig.~\ref{fig:mle_distribution_b}).
Using these parameters, we estimate that the total T-depth of the SFQG circuit multiplied by the required number of shots is $5.5 \times 10^7$, the end-to-end circuit would require 12k logical qubits, and we would need to execute T-gates at a code distance that can support $10^{8}$ logical operations.
Assuming the contract can be classically priced using Monte Carlo in 1 second to within the same error and confidence level \cite{chakrabarti2021threshold} and that the greeks can be calculated using a second-order finite-difference method applied to Monte Carlo pricings, the greeks of the basket option can be classically estimated in 8 seconds (four greeks, two pricings per greek).
Therefore, quantum advantage in calculating the greeks of this derivative contract would
require executing T-gates at a rate of 7MHz, $\sim 7$ times lower than the estimate of Chakrabarti et al. for quantum advantage in derivative pricing.
Moreover, because the quantum gradient circuit is sampled 60 times when we use the MLE method (see Fig.~\ref{fig:mle_distribution}), if we have parallel access to 60 QPUs on which the quantum circuit can be loaded and sampled simultaneously, we can achieve the same runtime as the serial execution if the logical cock rate of each device is $\sim 100$kHz, closer to current estimates of feasible logical clock rates around 10kHz \cite{Fowler2018}.

While the computational cost required for the numerical simulations of the
quantum gradient estimation algorithm limits our analysis to a maximum of $k=4$
greeks for practically relevant use cases, we expect that the algorithm can
scale favorably to derivative pricing problems of higher dimensionality,
motivated by the fact that most, if not all, derivative contracts of practical
interest have piecewise-linear payoffs~\cite{Hull}.
While Theorem~\ref{thm:smoothness_conditions}
provides a complexity of $\mathcal{O}(\sqrt{k}/\epsilon)$ for the class of
functions considered, it does not preclude higher-order speedups with respect
to $k$ for smoother functions.
For instance, the same gradient estimation algorithm applied to simple
polynomial functions can achieve a complexity of $\mathcal{O}(\log(k)/\epsilon)$ \cite{gilyen2019optimizing}.
Derivative contracts often include market parameters with little or no cross-dependence, i.e. $\partial^n f/(\partial x_1^m \partial x_2^{n-m})\sim 0$ for $n>1, m \in [1, n-1]$.
For example, the basket option from Sec.~\ref{sec:basket_option} without the knock-in feature satisfies $\partial^n V/(\partial S_i^n \partial S_j^{n-m})= 0$.
The absence of such higher-order terms allows us to pick values of $m$ and $l$ in the application of the GAW algorithm which lead to smaller overall oracular cost than what is required by Theorem \ref{thm:smoothness_conditions}.
An upper bound on the potential advantage compared to classical Monte Carlo is the case where the function is at most a second-degree polynomial in all $k$ variables, yielding an overall speedup of $\mathcal{O}(k)$.
Therefore, while we have made a first step in establishing the relevance and
promise of the quantum gradient estimation methods in the context of financial
derivative risk analysis, the possible extent of quantum speedup will be highly dependent on the nature of the price function for each derivative.

\section{Discussion}
We introduce a method to compute gradients of financial derivatives (greeks)
using the gradient algorithms from~\cite{Jordan_2005, gilyen2019optimizing}.
This method suggests additional quantum advantage is possible in risk analysis,
on top of the quadratic speedup of derivative
pricing~\cite{rebentrost2018quantum, Stamatopoulos_2020, chakrabarti2021threshold}.
Classically computing $k$ greeks with finite-difference methods - when
the underlying derivative is priced using Monte Carlo - has complexity
$\mathcal{O}(k/\epsilon^2)$ and straightforward extension of finite-difference
methods to derivative pricing using amplitude estimation provides a quadratic advantage with complexity $\mathcal{O}(k/\epsilon)$.
In this work, we explore an additional quadratic
advantage for overall complexity of $\mathcal{O}(\sqrt{k}/\epsilon)$.
The gradient estimation algorithm from~\cite{gilyen2019optimizing} guarantees
this quadratic advantage with respect to the number of greeks when the pricing
function satisfies the smoothness conditions of
Theorem~\ref{thm:smoothness_conditions}.
Because derivative pricing problems of practical interest in finance
involve numerous diverse multivariate price functions and generally have no
analytical solutions, understanding whether and which financial derivatives
satisfy the aforementioned smoothness conditions is a challenging task.
For this reason, we employ numerical methods to simulate the gradient
estimation algorithm for two example derivatives: a) a European call option
which has a closed-form solution and is used to establish the validity and
benchmarks of the algorithm and b) a path-dependent basket option which has no
known analytical solution and is representative of typical derivative price
functions.
We find that the quantum gradient algorithms not only succeed in estimating the
associated greeks for these examples with high probability, but that the
resulting query complexity is in fact significantly smaller than that
suggested by Theorem~\ref{thm:smoothness_conditions}
(Tables~\ref{tbl:numerical_vanilla} and \ref{tbl:numerical_basket}), suggesting
that the associated price functions are smoother than those studied in
Theorem~\ref{thm:smoothness_conditions}.

Another question we tackle in this work is the rigorous resource estimation of the
quantum oracles involved in the quantum gradient estimation algorithm.
Due to the extra cost associated with the
block-encoding and Hamiltonian simulation required to approximately construct the phase
oracle of Eq.~\eqref{eqn:general_phase_oracle} from the probability oracle of
Eq.~\eqref{eqn:A_operator} used in derivative pricing, in
Sec.~\ref{sec:second_order_exact_phase_oracle} we develop a method to construct a cheaper,
second-order ($m=1$) phase oracle exactly by taking advantage of the structure
of amplitude estimation.
An interesting question is whether this method can be extended to construct phase oracles for higher-order ($m > 1)$ gradient methods which would apply more generally to quantum gradient estimation problems.

In Sec.~\ref{sec:mle} we show that it is possible to enhance quantum gradient algorithms by employing maximum likelihood estimation (MLE), allowing us to determine the resources required to estimate gradients with precise confidence intervals and confidence levels.
Using this MLE method, in Sec.~\ref{sec:advantage_estimates} we estimate the resources required for quantum advantage in derivative market risk for typical use cases of practical interest.
We find that employing quantum gradient methods in derivative pricing lowers the logical clock rate estimate for quantum advantage from Chakrabarti et al.~\cite{chakrabarti2021threshold} by a factor of 7.

While finite-difference methods are still used in practice to
compute greeks, more recently classical automatic differentiation (AD) methods
have been gathering considerable interest because of their ability to
significantly reduce the associated computational costs, at the cost of
increased memory footprint~\cite{homescu2011adjoints, pages2016vibrato}.
In particular, the adjoint mode of automatic differentiation (AAD) in certain cases allows the computation of all $k$ gradients of a scalar function $f$ at a cost which is independent of $k$, meaning that the overall classical complexity cost in this case becomes $\mathcal{O}(\omega/\epsilon^2)$, for some constant $\omega$ depending on the function $f$ \cite{capriottiAD}.
While the complexity of the GAW quantum gradient estimation algorithm scales as $\mathcal{O}(\sqrt{k}/\epsilon)$ for the class of smooth functions in Theorem \ref{thm:smoothness_conditions}, in Sec.~\ref{sec:gaw_numerical} we saw that for practical use cases in finance, the algorithm scales 100x-200x times better than theoretical estimate from Theorem \ref{thm:smoothness_conditions} for a given $\epsilon$ (Tables \ref{tbl:numerical_vanilla} and \ref{tbl:numerical_basket}).
As such, depending on the practical scaling of the GAW algorithm to larger values of $k$ for finance use cases, it is possible that it could also outperform the complexity of AAD methods.
It is also interesting to consider whether a similar construct as that employed by AD can be applied in a quantum setting.
In Appendix \ref{app:auto_diff} we provide such a construct and show that in certain settings it can lead to similar performance profile as classical AD, in that the runtime of the algorithm is independent of the number of greeks at the expense of increased memory usage.
A detailed comparison of the performance between the quantum gradient algorithms and AD
methods is a worthy study on its own and it is left for future research.

Intuitively, the expensive part of quantum derivative pricing is
extracting the result through quantum amplitude estimation.
Therefore it is advantageous to
perform additional calculations involving that value before reading it out.
In this work we consider calculating gradients, but there are other associated
risk metrics that are equally of practical interest, such as the computation of
portfolio value-at-risk (VaR).
Based on our results, we are therefore
cautiously optimistic that further quantum advantage in the derivative pricing
subroutine can be leveraged at these higher levels of calculation and
aggregation, and suggests an additional research path going forward.

\begin{acknowledgments}
	We thank Rajiv Krishnakumar, Shouvanik Chakrabarti and Srinivasan
	Arunachalam for useful discussions regarding quantum gradient estimation
	algorithms, and Paul Burchard, Graham Griffiths, Alex Hurst, Dunstan Marris
	 and Elmer Tan for their technical and business insights regarding
	 financial derivatives and market risk.
\end{acknowledgments}

\bibliographystyle{apsrev4-1_custom}
\bibliography{greeks}   % name your BibTeX data base

\appendix

\section{Phase Kickback}
\label{app:phase_kickback}
In order for the Simulation-Free Quantum Gradient (SFQG) method described in Sec.~\ref{sec:sfqg} to work, care must be taken for the appropriate phase kickback to occur in Eq.~\eqref{eqn:Q_phase_kickback}, so that we get the correct gradient after the application of the inverse Quantum Fourier Transform.
Eq.~\eqref{eqn:Q_phase_kickback} includes the states $\ket{\Psi_{\pm}(\boldx)}$ which in general will interfere with the subsequent inverse Quantum Fourier Transform through their dependence on $\boldx$.
The application of the $\mathcal{Q}$ operator in the SFQG method creates the state

\begin{equation}
	\label{eqn:Q_phase_kickback_state}
 \ket{\Psi}=\frac{1}{\sqrt{N^k}}\sum_{\boldx}e^{2\pi
   	iS\theta(\boldx)}\ket{\boldx} \ket{\Psi_{+}(\boldx)} -
    	\frac{1}{\sqrt{N^k}}\sum_{\boldx}e^{-2\pi iS\theta(\boldx)
    	}\ket{\boldx}\ket{\Psi_{-}(\boldx)}.
\end{equation}
Applying the inverse Quantum Fourier Transform to the $\ket{\boldx}$ register, we get
\begin{equation}
 \frac{1}{{N^k}}\sum_{\boldx}\sum_{\boldy}e^{2\pi i (S\theta(\boldx) - \boldx \cdot \boldy/N)}\ket{\boldy} \ket{\Psi_{+}(\boldx)} -
    	\frac{1}{{N^k}}\sum_{\boldx}\sum_{\boldy}e^{-2\pi i (S\theta(\boldx) + \boldx \cdot \boldy/N)}\ket{\boldy}\ket{\Psi_{-}(\boldx)}.
\end{equation}
The probability of measuring a value $\ket{\boldz}$ in the first register is given by

\begin{align}
	\label{eqn:measurement_prob}
	\frac{1}{{N^{2k}}}& \left(\sum_{\boldxp}\sum_{\boldyp}e^{-2\pi i (S\theta(\boldxp) - \boldxp \cdot \boldyp/N)}\bra{\boldyp} \bra{\Psi_{+}(\boldxp)} -
    	\sum_{\boldxp}\sum_{\boldyp}e^{2\pi i (S\theta(\boldxp) + \boldxp \cdot \boldyp/N)}\bra{\boldyp}\bra{\Psi_{-}(\boldxp)} \right) (\ket{\boldz} \otimes \mathbb{I})(\bra{\boldz} \otimes \mathbb{I}) \nonumber \\
	&\left(\sum_{\boldx}\sum_{\boldy}e^{2\pi i (S\theta(\boldx) - \boldx \cdot \boldy/N)}\ket{\boldy} \ket{\Psi_{+}(\boldx)} -
    	\sum_{\boldx}\sum_{\boldy}e^{-2\pi i (S\theta(\boldx) + \boldx \cdot \boldy/N)}\ket{\boldy}\ket{\Psi_{-}(\boldx)} \right) \nonumber \\
  = \frac{1}{{N^{2k}}}&\Big(\sum_{\boldxp}\sum_{\boldx}e^{-2\pi i (S\theta(\boldxp) - \boldxp \cdot \boldz/N)}e^{2\pi i (S\theta(\boldx) - \boldx \cdot \boldz/N)}\braket{\Psi_{+}(\boldxp)|\Psi_{+}(\boldx)} \nonumber \\
 & + \sum_{\boldxp}\sum_{\boldx}e^{2\pi i (S\theta(\boldxp) + \boldxp \cdot \boldz/N)}e^{-2\pi i (S\theta(\boldx) + \boldx \cdot \boldz/N)}\braket{\Psi_{-}(\boldxp)|\Psi_{-}(\boldx)} \nonumber \\
 & - \sum_{\boldxp}\sum_{\boldx}e^{-2\pi i (S\theta(\boldxp) - \boldxp \cdot \boldz/N)}e^{-2\pi i (S\theta(\boldx) + \boldx \cdot \boldz/N)}\braket{\Psi_{+}(\boldxp)|\Psi_{-}(\boldx)} \nonumber \\
 & - \sum_{\boldxp}\sum_{\boldx}e^{2\pi i (S\theta(\boldxp) + \boldxp \cdot \boldz/N)}e^{-2\pi i (S\theta(\boldx) - \boldx \cdot \boldz/N)}\braket{\Psi_{-}(\boldxp)|\Psi_{+}(\boldx)}\Big)
\end{align}
In the derivative pricing context we consider in this manuscript, the $\mathcal{A}$ operator we consider implements the \emph{re-parameterization} method from \cite{chakrabarti2021threshold}.
In this case, the $\mathcal{A}$ operator can be written as the product of two operators $\mathcal{G}$ and $\mathcal{F}$.
The $\mathcal{G}$ operator loads standard normal distributions corresponding to the number of assets of the derivative contract and the timesteps used in the pricing

\begin{equation}
\mathcal{G} : \ket{0}_m \rightarrow \sum_i\sqrt{p_i}\ket{i},
\end{equation}
where the probabilities $p_i$ are independent of any market parameters.
Then, the $\mathcal{F}$ operator computes the payoff $g(i)$ using quantum arithmetic on $\ket{i}$, which is subsequently rotated into the amplitude of an ancilla qubit

\begin{equation}
	\mathcal{F} : \sum_i\sqrt{p_i}\ket{i}\ket{0}_q \rightarrow  \sum_i\sqrt{p_i}\ket{i}\ket{g(i)}(\sqrt{1-g(i)}\ket{0} + \sqrt{g(i)}\ket{1}).
\end{equation}
After the final rotation, the register $\ket{g(i)}$ can be uncomputed, so that the overall effect of the $\mathcal{A} = \mathcal{F}(\mathcal{G}\otimes \mathbb{I}^{\otimes q})$ operator can be written as

\begin{equation}
	\mathcal{A}\ket{0}_{m+1} = \sum_i\sqrt{p_i}\ket{i}(\sqrt{1-g(i)}\ket{0} + \sqrt{g(i)}\ket{1}),
\end{equation}
where $g(i) \in [0, 1]$.
When $\mathcal{A}$ is evaluated in superposition over a register $\ket{\boldx}$ representing tweaks to input market parameters as shown in Eq.\eqref{eqn:A_superposition}, the resulting state becomes

\begin{align}
	\label{eqn:A_operator_pricing}
	\mathcal{A}: \frac{1}{\sqrt{N^k}}\sum_{\boldx}\ket{\boldx}\ket{0}_{m+1} & \rightarrow \frac{1}{\sqrt{N^k}} \sum_{\boldx}\ket{\boldx}\sum_i\sqrt{p_i}\ket{i}(\sqrt{1-g(i, \boldx)}\ket{0} + \sqrt{g(i, \boldx)}\ket{1}) \nonumber \\
	& = \frac{1}{\sqrt{N^k}} \sum_{\boldx}\ket{\boldx}\left(e^{i\theta(\boldx)}\ket{\Psi_{+}(\boldx)} - e^{-i\theta(\boldx)}\ket{\Psi_{-}(\boldx)} \right),
\end{align}
with

\begin{equation}
	\label{eqn:psipm_states}
	\ket{\Psi_{\pm}(\boldx)}=\frac{1}{\sqrt{2}} \left( \sum_i \sqrt{p_i}\ket{i} \left(-i\sqrt{\frac{g(i, \boldx)}{\sum_i p_ig(i, \boldx)}}\ket{1} \pm \sqrt{\frac{1-g(i, \boldx)}{\sum_i p_i (1-g(i, \boldx))}}\ket{0} \right)  \right),
\end{equation}
and
\begin{equation}
	 \theta(\boldx) = \arcsin\left(\sqrt{\sum_i p_ig(i, \boldx)} \right)
\end{equation}
where $k$ is the dimension of the gradient we are estimating, and $N=2^n$, and $n$ qubits are used for the superposition $\sum_{\boldx}\ket{\boldx}$ in each dimension.
The gradient estimation algorithm requires that we choose $\boldx \ll 1$ so that the function $\theta(\boldx)$ is approximately linear in the vicinity of $\boldx=\mathbf{0}$.
In this regime, the $\ket{\Psi_{\pm}}$ states in Eq.\eqref{eqn:psipm_states} give $\braket{\Psi_{+}(\boldxp)|\Psi_{-}(\boldx)}=\braket{\Psi_{-}(\boldxp)|\Psi_{+}(\boldx)} \approx 0$ and $\braket{\Psi_{+}(\boldxp)|\Psi_{+}(\boldx)}=\braket{\Psi_{-}(\boldxp)|\Psi_{-}(\boldx)} \approx 1$.
The probability of measuring a value $\ket{z}$ in Eq.\eqref{eqn:measurement_prob} is then given by

\begin{equation}
	\frac{1}{{N^{2k}}}\sum_{\boldxp}\sum_{\boldx}e^{-2\pi i (S\theta(\boldxp) - \boldxp \cdot \boldz/N)}e^{2\pi i (S\theta(\boldx) - \boldx \cdot \boldz/N)} = \frac{1}{{N^{2k}}}\left | \sum_{\boldx}e^{2\pi i (S\theta(\boldx) - \boldx \cdot \boldz/N)}\right |^2,
\end{equation}
similarly to standard quantum phase estimation \cite{brassard2002quantum}.
Note that the application of the $\mathcal{Q}$ operator in Eq.~\eqref{eqn:Q_phase_kickback_state} induces the correct phase kickback to the $\ket{\boldx}$ register because the $\boldx$-dependence in Eq.\eqref{eqn:psipm_states} is limited to the amplitude of the last qubit.
If the probabilities $p_i$ become dependent on $\boldx$ or other qubit registers remain entangled with $\boldx$, the phase kickback fails.

\begin{figure}[th]
  \centering
  \includegraphics[width=0.85\linewidth]{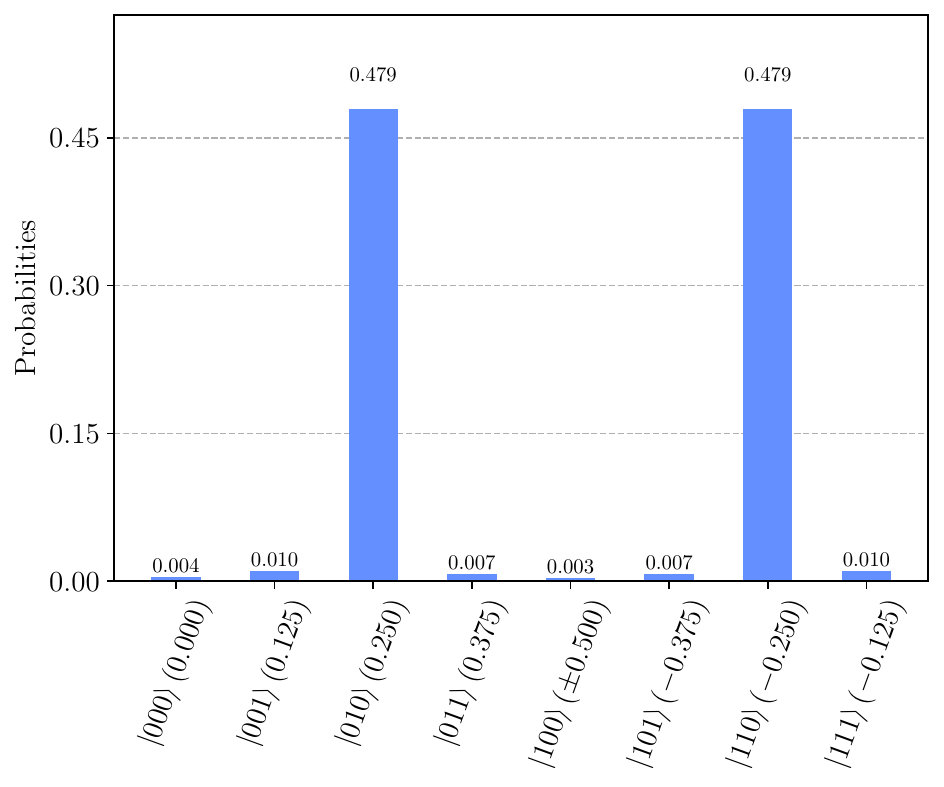}
  \caption{Simulated measurement outcomes after the application of Eq.~\eqref{eqn:Q_phase_kickback} and the subsequent inverse Quantum Fourier Transform when the $\mathcal{A}$ operator is in the form of Eq.~\eqref{eqn:A_operator_pricing}. The measurement outcomes correspond to estimates of the gradient $d\theta/dx$ at $x=0$ where $\theta = \arcsin(\sqrt{a})$ and $a = \sum_i{p_i}\sin^2(b*i+2x)$ for $b=0.405$ and $p=$[0.03149738, 0.08388914, 0.16118575, 0.22342773, 0.22342773, 0.16118575, 0.08388914, 0.03149738]. The probability peaks around $\pm 0.25$, which is the closest value representable with $n=3$ qubits to the expected $d\theta/dx \approx 0.253$.}
  \label{fig:phase_kickback}
\end{figure}

In Fig.~\ref{fig:phase_kickback} we show the simulated measurement outcomes after the creation of the state in Eq.~\eqref{eqn:Q_phase_kickback_state} and the subsequent inverse Quantum Fourier Transform for an example when the $\mathcal{A}$ operator is in the form of Eq.~\eqref{eqn:A_operator_pricing} and $k=1$.
The probabilities $p_i$ are taken from a normal distribution with unit variance defined on three qubits ($m=3$), normalized such that $\sum_{i}p_i=1$, and $g(i,x) = \sin^2(b*i + 2x)$ for $b=0.405$.
After applying the Quantum Fourier Transform and measuring the $\ket{x}$ register, we get an estimate of the gradient $d\theta/dx$ (or $-d\theta/dx$ as described by Eq.\eqref{eqn:measured_gradients}) at $x=0$ where $\theta = \arcsin(\sqrt{a})$ and $a = \sum_i{p_i}\sin^2(b*i+2x)$.
For the parameter values we have chosen, $d\theta/dx \approx 0.253$ at $x=0$.
We use three qubits for the register $\ket{x}$ ($n=3$) which allows us to resolve the gradient with accuracy $1/8=0.125$.
The value of $b$ was chosen for clarity, so that the resulting gradient is close to one of the values which can be represented exactly with three qubits.
We picked $l=\pi/256$ and $D=1$ and thus the $\mathcal{Q}$ operator is applied $S=N/Dl=2048$ times.
Because the $\mathcal{A}$ operator used in this simulation is in the form of Eq.~\eqref{eqn:A_operator_pricing}, the correct phase is kicked back to the $\ket{x}$ register, and the inverse Quantum Fourier Transform generates two probability peaks around $\pm d\theta/dx$ where $d\theta/dx \approx 0.253$

\section{Automatic Differentiation and Multi-Objective QAE}
\label{app:auto_diff}

Another way to compute gradients in the considered setting is automatic differentiation (AD) \cite{homescu2011adjoints, pages2016vibrato, capriottiAD}.
AD repeatedly applies the chain rule to every elementary arithmetic operation that is used to compute an objective function and keeps track of the analytical gradient throughout the calculation.
Different variants of AD exist \cite{homescu2011adjoints} and it has been shown that in many practical applications the gradient can be computed at only a constant overhead, independent of the dimension \cite{capriottiAD}.
In some cases the overhead can even be stated to be bounded by a factor of four compared to evaluating the function alone at the expense of larger memory requirements \cite{Capriotti2010FastCG}.

Suppose now we want to estimate an expectation value $\mathbb{E}(g(S, x))$ for a payoff function $g$, a random variable $S$ and some given parameters $x$ as well as the corresponding gradient $\nabla_x \mathbb{E}(g(S, x))$.
To construct the probability oracle $\mathcal{A}(x)$ required by QAE for a fixed $x$, we usually first create a weighted superposition of all scenarios, then evaluate the corresponding payoff for each scenario, and last prepare an objective qubit, i.e., we get
\begin{eqnarray}
\sum_{j=0}^{2^m-1} \sqrt{p_j} \ket{s_j}\ket{g(s_j, x)}\left( \sqrt{1 - g(s_j, x)}\ket{0} + \sqrt{g(s_j, x)}\ket{1} \right),
\end{eqnarray}
such that the probability of measuring $\ket{1}$ in the last qubit corresponds to $\mathbb{E}(g(S, x))$, and where the $s_j$ denote the possible realizations of $S$ represented by $m$ qubits and the $p_j$ denote the corresponding probabilities.

For every scenario $\ket{s_j}$, we apply quantum arithmetic to compute the payoff $\ket{g(s_j, x)}$.
Thus, for each $s_j$, we can also use AD in the same way as classically to compute the gradient $\nabla_x g(s_j, x)$, while using at most twice the resources required classically due to the need of a reversible implementation \cite{bennett_reversible_computation}.
Thus, with a constant overhead compared to the evaluation of the expectation value, this results in the state
\begin{eqnarray}
\sum_{j=0}^{2^m-1} \sqrt{p_j} \ket{s_j}\ket{g(s_j, x)}\bigotimes_{i=1}^k \ket{\partial_i g(s_j, x)}.
\label{eq:ad_qae_gradient}
\end{eqnarray}
In the following, we show how to use QAE to read out multiple objectives defined on the same random variables, which then immediately applies to the gradient as constructed in Eq.~\eqref{eq:ad_qae_gradient}.

Suppose a random variable $S$ and a set of functions $f_i$, $i=1, \ldots, k$, that map realizations of $S$ to $\mathbb{R}$.
Further,  suppose we are interested in estimating the expectation values $\mathbb{E}(f_i(S))$ for all $i$, and that we can construct a state of the form
\begin{eqnarray}
\sum_{j=0}^{2^m-1} \sqrt{p_j} \ket{s_j}\bigotimes_{i=1}^k \ket{f_i(s_j)}.
\end{eqnarray}
Then, to estimate the values $\mathbb{E}(f_i(S))$, we first introduce $k$ additional $m$-qubit registers $\ket{c_i}$, each initialized with some value $c_i$, and second, we use quantum arithmetic to compute the sum
\begin{eqnarray}
\sum_{i=1}^{k} c_i f_i(s_j)
\label{eq:ad_qae_objective}
\end{eqnarray}
into another register.
In other words, we construct an operator that acts as
\begin{eqnarray}
\bigotimes_{i=1}^k \ket{c_i} \ket{0}\bigotimes_{i=1}^k \ket{0} \ket{0} \mapsto
\bigotimes_{i=1}^k \ket{c_i} \sum_{j=0}^{2^m-1} \sqrt{p_j} \ket{s_j}\bigotimes_{i=1}^k \ket{f_i(s_j)} \ket{\sum_{i=1}^k c_i f_i(s_j)}.
\end{eqnarray}
By adding an objective qubit and applying a rotation controlled by the last register we can also use this to construct a probability oracle $\mathcal{A}(c)$ that corresponds to the function
\begin{eqnarray}
f(c) &=& \mathbb{E} \left( \sum_{i=1}^k c_i f_i(S)  \right).
\end{eqnarray}
This is a linear function in $c$ and using the quantum gradient algorithm with respect to $c$ results in
\begin{eqnarray}
\nabla_c f(c) = \left(\mathbb{E}( f_1(S)), \ldots, \mathbb{E}( f_k(S)) \right)^T,
\end{eqnarray}
i.e., in the read out of all $k$ expectation values.

If the values of $f_i$ are such that the weighted sum in Eq.~\eqref{eq:ad_qae_objective} satisfies $\sum_{i=1}^{k} c_i f_i(s_j) \le 1$, since the function by construction is linear in $c$, the resulting complexity of the quantum gradient algorithm for a target accuracy $\epsilon > 0$ scales as $\mathcal{\tilde O}(1 / \epsilon)$, i.e., independent of $k$ (ignoring logarithmic terms),  following \cite[Thm. ~23, arxiv version]{gilyen2019optimizing}.
The multi-objective QAE requires $k \cdot m$ additional qubits as well as the weighted sum in Eq.~\eqref{eq:ad_qae_objective}, which can be computed in logarithmic depth by using a divide-and-conquer summation scheme.
If on the other hand, Eq.~\eqref{eq:ad_qae_objective} needs to be normalized by dividing the weighted sum with a factor $D$, the complexity of the algorithm becomes $\mathcal{\tilde O}(1 / \epsilon D)$.
When $f_i \sim 1, \forall i$, we would need to choose $D \sim 1/k$, which adds a factor of $k$ back to the complexity of the algorithm, negating the advantage of this method.

Since Eq.~\eqref{eq:ad_qae_gradient} has the required shape, we can immediately apply the multi-objective QAE to evaluate the gradient that has been evaluated using AD implemented by quantum arithmetic.
If no additional normalization is required, we can combine AD and (multi-objective) QAE to get the gradient algorithm with runtime $\mathcal{\tilde O}(1/\epsilon)$, i.e.,  independent of the dimension and with a quadratic speed-up in the accuracy.
Thus, like classically, AD could represent a promising approach to estimate market risks with a significant advantage over finite difference schemes.
What remains to be analyzed in more depth, is how this affects the required memory, i.e. qubits, and how to best automate automatic differentiation for reversible quantum arithmetic.

\end{document}